\documentstyle{livrev}

\RequirePackage{epsf}
\RequirePackage{longtable}

\begin{document}

\title{Testing General Relativity with Pulsar Timing}

\author{Ingrid H. Stairs \\
        Dept. of Physics and Astronomy \\
        University of British Columbia \\
        6224 Agricultural Road \\
        Vancouver, B.C. \\
        V6T 1Z1 Canada\\
        e-mail:stairs@astro.ubc.ca \\
\\
\small{(last modified: June 9, 2003)}
}

\date{}
\maketitle

\begin{abstract}
Pulsars of very different types -- isolated objects, and binaries with
short- and long-period orbits, white-dwarf and neutron-star companions
-- provide the means to test both the predictions of general
relativity and the viability of alternate theories of gravity.  This
article presents an overview of pulsars, then discusses the current
status and future prospects of tests of equivalence principle
violations and strong-field gravitational experiments.
\end{abstract}


\newpage


\section{Introduction}
\label{section:introduction}

Since their discovery in 1967 \cite{hbp+68}, radio pulsars have
provided insights into physics on length scales covering the range
from 1\,m (giant pulses from the Crab pulsar \cite{hkwe03}) to
10\,km (neutron star) to kpc (Galactic) to hundreds of Mpc
(cosmological).  Pulsars present an extreme stellar environment, with
matter at nuclear densities, magnetic fields of 10$^8$\,G to nearly
10$^{14}$\,G, and spin periods ranging from 1.5\,ms to 8.5\,s.  The
regular pulses received from a pulsar each correspond to a single
rotation of the neutron star.  It is by measuring the deviations from
perfect observed regularity that information can be derived about the
neutron star itself, the interstellar medium between it and the Earth,
and effects due to gravitational interaction with binary companion
stars.

In particular, pulsars have proved to be remarkably successful
laboratories for tests of the predictions of general relativity
(GR). The tests of GR that are possible through pulsar timing fall
into two broad categories: setting limits on the magnitudes of
parameters that describe violation of equivalence principles, often
using an ensemble of pulsars, and verifying that the measured
post-Keplerian timing parameters of a given binary system match the
predictions of strong-field GR better than those of other theories.
Long-term millisecond pulsar timing can also be used to set limits on
the stochastic gravitational-wave background (e.g.,
\cite{ktr94,lom01,jb03}), as can limits on orbital variability in
binary pulsars for even lower wave frequencies (e.g.,
\cite{bcr83,kop97}).  However these are not tests of the same type of precise
prediction of GR and will not be discussed here.  This review will
present a brief overview of the properties of pulsars and the
mechanics of deriving timing models, and will then proceed to describe
the various types of tests of GR made possible by both single and
binary pulsars.


\newpage
\section{Pulsars, Observations and Timing}
\label{section:pulsar_intro}

The properties and demographics of pulsars, as well as pulsar search
and timing techniques, are thoroughly covered in the article by
Lorimer in this series \cite{lor01}.  This section will present only
an overview of the topics most important to understanding the
application of pulsar observations to tests of GR.

\newpage
\subsection{Pulsar Properties}
\label{section:pulsars}

Radio pulsars were firmly established as being neutron stars with the
discovery of the pulsar in the Crab nebula \cite{sr68}; its 33-ms
period was too fast to permit a pulsating or rotating white dwarf,
leaving a rotating neutron star as the only surviving model
\cite{pac68,gol68}.  The 1982 discovery of a 1.5-ms pulsar,
PSR~B1937+21 \cite{bkh+82}, led to the realization that, in addition
to the ``young'' Crab-like pulsars born in recent supernovae, there
exists a separate class of older ``millisecond'' or ``recycled''
pulsars, which have been spun up to faster periods by accretion of
matter and angular momentum from an evolving companion star.  See, for
example, \cite{bv91} and
\cite{pk94} for reviews of the evolution of such binary systems. It is
precisely these recycled pulsars that form the most valuable resource
for tests of GR.

\begin{figure}[h]
  \def\epsfsize#1#2{0.4#1} 
\centerline{\epsfbox{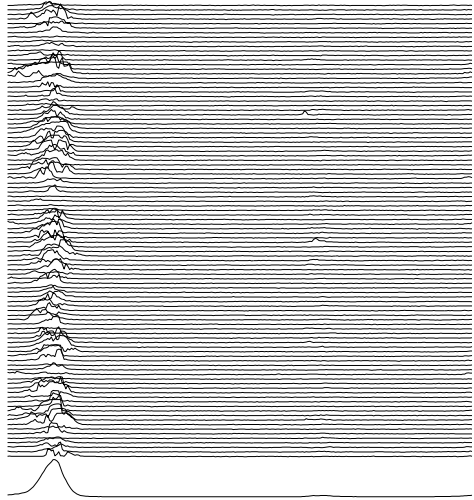}}
  \caption{\it Top: 100 single pulses from the 253-ms pulsar B0950+08, demonstrating
  pulse-to-pulse variability in shape and intensity.  Bottom:
  Cumulative profile for this pulsar over 5 minutes (about 1200
  pulses); this approaches the reproducible standard profile.
  Observations taken with the Green Bank Telescope \cite{gbt}.
  Stairs, unpublished.}  \label{figure:indiv}
\end{figure}

The exact mechanism by which a pulsar radiates the energy observed as
radio pulses is still a subject of vigorous debate.  The basic
picture of a misaligned magnetic dipole, with coherent radiation from
charged particles accelerated along the open field lines above the
polar cap \cite{gj69,stu71}, will serve adequately for the purposes of this
article, in which pulsars are treated as a tool to probe other
physics.  While individual pulses fluctuate severely in both intensity
and shape (see Fig.~\ref{figure:indiv}), a profile ``integrated'' over
several hundred or thousand pulses (i.e., a few minutes) yields a
shape -- a ``standard profile'' -- that is reproducible for a given
pulsar at a given frequency.  (There is generally some evolution of
pulse profiles with frequency, but this can usually be taken into
account.)  It is the reproducibility of time-averaged profiles that
permits high-precision timing.

Of some importance later in this article will be models of the pulse
beam {\it shape}, the envelope function which forms the standard
profile.  The collection of pulse profile shapes and polarization
properties have been used to formulate phenomenological descriptions
of the pulse emission regions.  At the simplest level (e.g.,
reference~\cite{ran83} and other papers in that series), the
classifications can be broken down into Gaussian-shaped ``core''
regions with little linear polarization and some circular
polarization, and double-peaked ``cone'' regions with stronger linear
polarization and S-shaped position angle swings in accordance with the
``Rotating Vector Model'' (RVM; see \cite{rc69a}).  While these models
prove helpful for evaluating observed changes in the profiles of
pulsars undergoing geodetic precession, there are ongoing disputes in
the literature as to whether the core/cone split is physically
meaningful, or whether both types of emission are simply due to patchy
strength of a single emission region (e.g., \cite{lm88}).

\newpage
\subsection{Pulsar Observations}
\label{section:observing}

\begin{figure}[h]
  \def\epsfsize#1#2{0.4#1} 
\centerline{\epsfbox{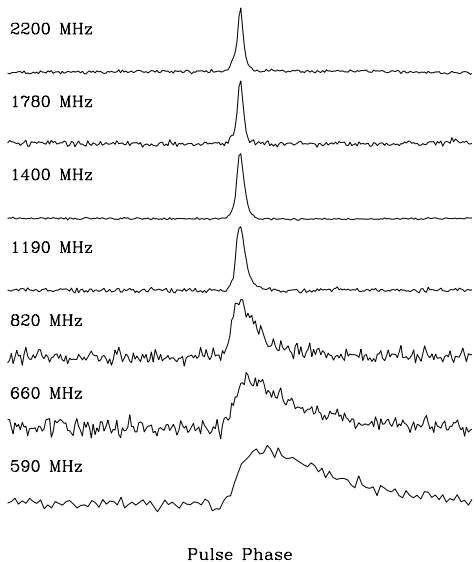}}
  \caption{\it Pulse profile shapes for PSR~J1740$-$3052 at multiple
  frequencies, aligned by pulse timing.  The full pulse period is
  displayed at each frequency.  The growth of an exponential scattering
  tail at low frequencies is evident.  All observations taken with the
  Green Bank Telescope \cite{gbt} (Stairs, unpublished), except for
  the 660-MHz profile which was acquired at the Parkes telescope
  \cite{parkes,sml+01}.}  \label{figure:1740scatt}
\end{figure}

A short description of pulsar observing techniques is in order.  As
pulsars have quite steep radio spectra (e.g.,\cite{mkkw00a}), they are
strongest at frequencies $f_0$ of a few hundred MHz.  At
these frequencies, the propagation of the radio wave through the
ionized interstellar medium (ISM) can have quite serious effects on
the observed pulse.  Multipath scattering will cause the profile to be
convolved with an exponential tail, blurring the sharp profile edges
needed for the best timing.  Figure~\ref{figure:1740scatt} shows an
example of scattering; the effect decreases with sky frequency as
roughly $f_0^{-4}$ (e.g., \cite{mt77}) and thus affects timing
precision less at higher observing frequencies.  A related effect is
scintillation: interference between the rays traveling along the
different paths causes time- and frequency-dependent peaks and valleys
in the pulsar's signal strength.  The decorrelation bandwidth, across
which the signal is observed to have roughly equal strength, is
related to the scattering time and scales as $f_0^{4}$ (e.g.,
\cite{mt77}).  There is little any instrument can do to compensate for
these effects; wide observing bandwidths at relatively high
frequencies and generous observing time allocations are the only ways
to combat these problems.

\begin{figure}[h]
  \def\epsfsize#1#2{0.4#1} 
\centerline{\epsfbox{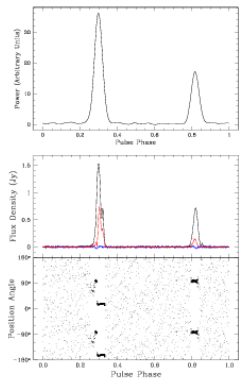}}
  \caption{\it Pulse profile of the fastest rotating pulsar,
  PSR~B1937+21, observed with the 76-m Lovell telescope at Jodrell
  Bank Observatory \cite{jbo}.  The top panel shows the
  total-intensity profile derived from a filterbank observation (see
  text); the true profile shape is convolved with the response of the
  channel filters.  The lower panel shows the full-Stokes observation
  with a coherent dedispersion instrument \cite{stc99,sst+00}.  Total
  intensity is indicated by black lines, and linear and circular power
  by red and blue lines respectively.  The position angle of the
  linear polarization is plotted twice.  The coherent dedispersion
  observation results in a much sharper and more detailed pulse
  profile, less contaminated by instrumental effects and more closely
  resembling the pulse emitted by the rotating neutron star.  Much
  better timing precision can be obtained with these sharper pulses.}
  \label{figure:cdedisp}
\end{figure}

Another important effect induced by the ISM is the dispersion of the
traveling pulses.  Acting as a tenuous electron plasma, the ISM causes
the wavenumber of a propagating wave to become frequency-dependent.
By calculating the group velocity of each frequency component, it is
easy to show (e.g.,\cite{mt77}) that lower frequencies will arrive at
the telescope later in time than the higher-frequency components,
following a $1/f^2$ law.  The magnitude of the delay is completely
characterized by the dispersion measure (DM), the integrated electron
content along the line of sight between the pulsar and the Earth.  All
low-frequency pulsar observing instrumentation is required to address
this dispersion problem if the goal is to obtain profiles suitable for
timing.  One standard approach is to split the observing bandpass into
a multichannel ``filterbank,'' to detect the signal in each channel,
and then to realign the channels following the $1/f^2$ law when
integrating the pulse.  This method is certainly adequate for slow
pulsars and often for nearby millisecond pulsars.  However, when the
ratio of the pulse period to its DM becomes small, much sharper
profiles can be obtained by sampling the voltage signals from the
telescope prior to detection, and convolving the resulting time series
with the inverse of the easily calculated frequency-dependent filter
imposed by the ISM.  As a result, the pulse profile is perfectly
aligned in frequency, without any residual dispersive smearing caused
by finite channel bandwidths.  As well, full-Stokes information can be
obtained without significant increase in analysis time, allowing
accurate polarization plots to be easily derived.  This ``coherent
dedispersion'' technique \cite{hr75} is now in widespread use across
normal observing bandwidths of several tens of MHz, thanks to the
availability of inexpensive fast computing power (e.g.,
\cite{bdz+97,jcpu97,sst+00}).  Some of the highest-precision
experiments described below have used this approach to collect their
data.  Figure~\ref{figure:cdedisp} illustrates the advantages of this
technique.

\newpage
\subsection{Pulsar Timing}
\label{section:timing_howto}

Once dispersion has been removed, the resultant time series is
typically folded modulo the expected pulse period, in order to build
up the signal strength over several minutes and to obtain a stable
time-averaged profile.  The pulse period may not be very easily
predicted from the discovery period, especially if the pulsar happens
to be in a binary system.  The goal of pulsar timing is to develop a
model of the pulse phase as a function of time, so that all future
pulse arrival times can be predicted with a good degree of accuracy.

The profile accumulated over several minutes is compared by
cross-correlation with the ``standard profile'' for the pulsar at that
observing frequency.  A particularly efficient version of the
cross-correlation algorithm compares the two profiles in the frequency
domain \cite{tay92}.  Once the phase shift of the observed profile
relative to the standard profile is known, that offset is added to the
start time of the observation in order to yield a ``Time of Arrival''
(TOA) that is representative of that few-minute integration.  In
practice, observers frequently use a time- and phase-stamp near the
middle of the integration in order to minimize systematic errors due
to a poorly known pulse period.  As a rule, pulse timing precision is
best for bright pulsars with short spin periods, narrow profiles with
steep edges, and little if any profile corruption due to interstellar
scattering.

With a collection of TOAs in hand, it becomes possible to fit a model
of the pulsar's timing behaviour, accounting for every rotation of the
neutron star.  Based on the magnetic dipole model \cite{pac68,gol68},
the pulsar is expected to lose rotational energy and thus ``spin
down''.  The primary component of the timing model is therefore a
Taylor expansion of the pulse phase $\phi$ with time $t$:
\begin{equation}
\phi = \phi_0 + \nu(t-t_0) + \frac{1}{2}\dot\nu(t-t_0)^2 +\ldots 
\label{equation:phase}
\end{equation}
where $\phi_0$ and $t_0$ are a reference phase and time, respectively,
and the pulse frequency $\nu$ is the time derivative of the pulse
phase.  Note that the fitted parameters $\nu$ and $\dot \nu$ and the
magnetic dipole model can be used to derive an estimate of the surface
magnetic field $B\,\sin\alpha$:
\begin{equation}\label{equation:Bfield}
B\,\sin\alpha =\, \left(\frac{-3I\dot\nu c^3}{8\pi^2R^6\nu^3}\right)^{1/2}\,
\approx\,
3.2\times10^{19}\left(\frac{-\dot\nu}{\nu^3}\right)^{1/2}\,G,
\end{equation}
where $\alpha$ is the inclination angle between the pulsar spin axis
and the magnetic dipole axis, $R$ is the radius of the neutron star,
about 10$^6$\,cm, and the moment of inertia is
$I\,\simeq\,10^{45}$\,g\,cm$^2$.  In turn, integration of the energy
loss, along with the assumption that the pulsar was born with infinite
spin frequency, yields a ``characteristic age'' $\tau_{\rm c}$ for the
pulsar:
\begin{equation}\label{equation:tauc}
\tau_{\rm c}\, = \, -\frac{\nu}{2\dot\nu}.
\end{equation}

\newpage
\subsubsection{Basic Transformation}
\label{section:timing_basic}

Equation~\ref{equation:phase} refers to pulse frequencies and times in
a reference frame that is inertial relative to the pulsar.  TOAs
derived in the rest frame of a telescope on the Earth must therefore
be translated to such a reference frame before
Equation~\ref{equation:phase} can be applied.  The best approximation
available for an inertial reference frame is that of the Solar System
Barycentre (SSB).  Even this is not perfect; many of the tests of GR
described below require correcting for the small relative
accelerations of the SSB and the centre-of-mass frames of binary
pulsar systems.  But certainly for the majority of pulsars it is
adequate.  The required transformation between a TOA at the telescope,
$\tau$, and the emission time $t$ from the pulsar is:
\begin{equation}
t =  \tau-D/f^2 + \Delta_{R\odot} + \Delta_{E\odot}
  -\Delta_{S\odot} - \Delta_R - \Delta_E - \Delta_S\,.
\label{equation:orbit}
\end{equation}
Here $D/f^2$ accounts for the dispersive delay in seconds of the
observed pulse relative to infinite frequency; the parameter $D$ is
derived from the pulsar's dispersion measure by $D={\rm
DM}/2.41\times10^{-4}$\, Hz, with DM in units of pc\,cm$^{-3}$ and the
observing frequency $f$ in MHz.  The Roemer term, $\Delta_{R\odot}$,
takes out the travel time across the Solar System based on the
relative positions of the pulsar and the telescope, including, if
needed, the proper motion and parallax of the pulsar.  The Einstein
delay, $\Delta_{E\odot}$, accounts for the time dilation and
gravitational redshift due to the Sun and other masses in the Solar
System, while the Shapiro delay, $\Delta_{S\odot}$, expresses the
excess delay to the pulsar signal as it travels through the
gravitational well of the Sun --- a maximum delay of about 120\,$\mu$s
at the limb of the Sun; see \cite{bh86} for a fuller discussion of
these terms.  The terms $\Delta_R$, $\Delta_E$, $\Delta_S$ in
Equation~\ref{equation:orbit} account for similar ``Roemer'',
``Einstein'' and ``Shapiro'' delays within the pulsar binary system,
if needed, and will be discussed in Section~\ref{section:binary}
below.  Most observers accomplish the model fitting, accounting for
these delay terms, using the program {\sc tempo} \cite{tempo}.  The
correction of TOAs to the reference frame of the SSB requires an
accurate ephemeris for the Solar System.  The most commonly used
ephemeris is the ``DE200'' standard from the Jet Propulsion Laboratory
\cite{sta82}.  It is also clear that accurate time-keeping is of
primary importance in pulsar modeling.  General practice is to derive
the time-stamp on each observation from the Observatory's local time
standard --- typically a Hydrogen maser --- and to apply, retroactively,
corrections to well-maintained time standards such as UTC(BIPM),
Universal Coordinated Time as maintained by the Bureau International
des Poids et Mesures in Paris.

\newpage
\subsubsection{Binary Pulsars}
\label{section:binary}

The terms $\Delta_R$, $\Delta_E$, $\Delta_S$ in
Equation~\ref{equation:orbit} describe the ``Roemer'', ``Einstein''
and ``Shapiro'' delays within a pulsar binary system.  The majority of
binary pulsar orbits are adequately described by five Keplerian
parameters: the orbital period $P_{\rm b}$, the projected semi-major
axis $x$, the eccentricity $e$, and the longitude $\omega$ and epoch
$T_0$ of periastron.  The angle $\omega$ is measured from the line of
nodes, ${\bf \Omega}$, where the pulsar orbit intersects the plane of
the sky.  In many cases, one or more relativistic corrections to the
Keplerian parameters must also be fit.  Early relativistic timing
models, developed in the first years after the discovery of
PSR~B1913+16, either did not provide a full description of the orbit
(e.g., \cite{bt76}), or else did not define the timing parameters in a way
that allowed deviations from GR to be easily identified (e.g.,
\cite{eps77,hau85}).  The best modern timing model
\cite{dd86,tw89,dt92}) incorporates a number of ``post-Keplerian''
timing parameters which are included in the description of the three
delay terms, and which can be fit in a completely phenomenological
manner.  The delays are defined primarily in terms of the phase of the
orbit, defined by the eccentric anomaly $u$ and true anomaly $A_e(u)$,
as well as $\omega$, $P_{\rm b}$ and their possible time
derivatives.  These are related by:
\begin{equation}
u-e\sin u = 2\pi \left[ \left( {{T-T_0}\over{P_{\rm b}}} \right) -
  {{\dot P_{\rm b}}\over 2} \left( {{T-T_0}\over{P_{\rm b}}} \right)^2
  \right], \label{eqn:eccen_anom} 
\end{equation}
\begin{equation} 
A_e(u) = 2 \arctan \left[ \left( {{1+e}\over{1-e}}
  \right)^{1/2} \tan {u\over2} \right], \label{eqn:true_anom} 
\end{equation}
\begin{equation} 
\omega = \omega_0 + \left(\frac{P_{\rm b}\,\dot{\omega}}{2\pi}\right) 
         A_e(u),  \label{eqn:omega}
\end{equation}
where $\omega_0$ is the reference value of $\omega$ at time $T_0$.
The delay terms then become:
\begin{equation} 
\Delta_R = x \sin\omega (\cos u -e(1+\delta_r)) + x
(1-e^2(1+\delta_{\theta})^2)^{1/2}\cos\omega \sin u,
\label{equation:deltaR}
\end{equation}
\begin{equation} 
\Delta_E = \gamma \sin u, \label{equation:deltaE} 
\end{equation}
\begin{equation} 
\Delta_S = -2r \ln \left\{ 1-e\cos u - s \left[
  \sin\omega (\cos u - e) + (1-e^2)^{1/2} \cos\omega \sin u
  \right] \right\}.\, \label{equation:deltaS} 
\end{equation}
Here $\gamma$ represents the combined time dilation and gravitational
redshift due to the pulsar's orbit, and $r$ and $s$ are, respectively,
the range and shape of the Shapiro delay.  Together with the orbital
period derivative $\dot P_{\rm b}$ and the advance of periastron $\dot
\omega$, they make up the post-Keplerian timing parameters that can be
fit for various pulsar binaries.  A fuller description of the timing
model also includes the aberration parameters $\delta_r$ and
$\delta_{\theta}$, but these parameters are not in general separately
measurable.  The interpretation of the measured post-Keplerian timing
parameters will be discussed in the context of double-neutron-star
tests of GR in Section~\ref{section:strong_field}.

\newpage
\section{Tests of GR --- Equivalence Principle Violations}
\label{section:sep}

Equivalence principles are fundamental to gravitational theory; for
full descriptions see e.g., \cite{mtw73} or \cite{wil93}.  Newton
formulated what may be considered the earliest such principle, now
called the ``Weak Equivalence Principle'' (WEP).  It states that in an
external gravitational field, objects of different compositions and
masses will experience the same acceleration.  The Einstein
Equivalence Principle (EEP) includes this concept as well as those of
Lorentz invariance (non-existence of preferred reference frames) and
positional invariance (non-existence of preferred locations) for
non-gravitational experiments.  This principle leads directly to the
conclusion that non-gravitational experiments will have the same
outcomes in inertial and in freely-falling reference frames.  The
Strong Equivalence Principle (SEP) adds Lorentz and positional
invariance for gravitational experiments, thus including experiments
on objects with strong self-gravitation.  As GR incorporates the SEP,
and other theories of gravity may violate all or parts of it, it is
useful to define a formalism which allows immediate identifications of
such violations.

The parametrized post-Newtonian (PPN) formalism was developed
\cite{wn72} to provide a uniform description of the
weak-gravitational-field limit, and to facilitate comparisons of rival
theories in this limit.  This formalism requires 10 parameters
($\gamma_{\rm PPN}$, $\beta$, $\xi$,$\alpha_1$, $\alpha_2$,
$\alpha_3$, $\zeta_1$, $\zeta_2$, $\zeta_3$ and $\zeta_4$) which are
fully described in the article by Will in this series \cite{wil01}
and whose physical meanings are nicely summarized in Table~2 of that
article.  (Note that $\gamma_{\rm PPN}$ is not the same as the
Post-Keplerian pulsar timing parameter $\gamma$.)  Damour and
Esposito-Far\`{e}se \cite{de92,de96a} extended this formalism to
include strong-field effects for generalized tensor-multiscalar
gravitational theories.  This allows a better understanding of limits
imposed by systems including pulsars and white dwarfs, for which the
amounts of self-gravitation are very different.  Here, for instance,
$\alpha_1$ becomes $\hat {\alpha_1} = \alpha_1 +
\alpha_1^{\prime}(c_1+c_2)+...$, where $c_i$ describes the
``compactness'' of mass $m_i$ . The compactness can be written $c_i =
-2\partial \ln m_i/\partial \ln G
\simeq -2(E^{\rm grav}/(m c^2))_{i}$, where $G$ is Newton's constant and 
$E^{\rm grav}_i$ is the gravitational self-energy of mass $m_i$, about
$-0.2$ for a neutron star (NS) and $-10^{-4}$ for a white dwarf (WD).
Pulsar timing has the ability to set limits on $\hat {\alpha_1}$,
which tests for the existence of preferred-frame effects (violations
of Lorentz invariance); $\hat \alpha_3$, which, in addition to testing for
preferred-frame effects, also implies non-conservation of momentum if
non-zero; and $\zeta_2$, which is also a non-conservative
parameter. Pulsars can also be used to set limits on other
SEP-violation effects which constrain combinations of the PPN
parameters: the Nordtvedt (``gravitational Stark'') effect, dipolar
gravitational radiation, and variation of Newton's constant.  The next
sections will discuss the current pulsar timing limits on each of
these effects.  Table~\ref{table:PPN_params} summarizes the PPN and
testable parameters, giving the best pulsar and solar-system limits.

{\tiny
\begin{table}[hp]
\begin{center}
\begin{tabular} {l p{1.5cm} p{1.5cm} l p{1.5cm} l}
\hline
Parameter & Physical Meaning & Solar-system test & Limit & Pulsar test & Limit \\
\hline\hline
$\gamma_{\rm PPN}$ & Space curvature produced by unit rest mass & VLBI, light deflection;\linebreak  measures\linebreak $|\gamma_{\rm PPN}\!-\!1|$ & $3\times 10^{-4}$  & & \\
$\beta$ & Non-linearity in superposition law for gravity & Perihelion shift of Mercury;\linebreak measures \linebreak $|\beta\!-\!1|$ & $3\times 10^{-3}$ & & \\ 
$\xi$ & Preferred-location effects & Solar alignment with ecliptic  & $4\times10^{-7}$  & & \\
$\alpha_1$ & Preferred-frame effects & Lunar laser ranging & $10^{-4}$ & Ensemble of binary pulsars  & $1.4 \times 10^{-4}$ \\
$\alpha_2$ & Preferred-frame effects & Solar alignment with ecliptic  & $4\times10^{-7}$ & & \\
$\alpha_3$ & Preferred-frame effects and non-conservation of momentum & Perihelion shift of Earth and Mercury & $2\times10^{-7}$ & Ensemble of binary pulsars  & $1.5\times10^{-19}$\\
$\zeta_1$ & Non-conservation of momentum & Combined PPN limits & $2\times10^{-2}$ & & \\
$\zeta_2$ & Non-conservation of momentum & & & Limit on $\ddot P$ for PSR~B1913+16  & $ 4\times10^{-5}$ \\
$\zeta_3$ & Non-conservation of momentum & Lunar acceleration & $10^{-8}$ & & \\
$\zeta_4$ & Non-conservation of momentum & Not independent & & & \\
$\eta$, $\Delta_{\rm net}$ & Gravitational Stark effect & Lunar laser ranging  & $10^{-3}$ & Ensemble of binary pulsars  & $9\times10^{-3}$ \\
$(\alpha_{c_1} - \alpha_0)^2$ & Pulsar coupling to scalar field & & & Dipolar gravitational radiation for PSR~B0655+64 & $2.7\times10^{-4}$\\
$\dot G/G$ & Variation of Newton's constant & Laser ranging to the Moon and Mars  & $6\times 10^{-12}$\,yr$^{-1}$ & Changes in Chandrasekhar mass  & $4.8 \times 10^{-12}$\,yr$^{-1}$ \\
\hline
\end{tabular}
\caption{PPN and other testable parameters, with the best 
solar-system and binary pulsar tests.  Physical meanings and most of
the solar-system references are taken from the compilations by Will
\cite{wil01}.  References: $\gamma_{\rm PPN}$, Solar-system: \cite{ema+99}; 
$\beta$, Solar-system: \cite{sha90b}; $\xi$, Solar-system: \cite{nor87}; 
$\alpha_1$, Solar-system: \cite{mnv96}, pulsar: \cite{wex00}; 
$\alpha_2$, Solar-system: \cite{nor87,wil93}; 
$\alpha_3$, Solar-system: \cite{wil93}, pulsar: \cite{wex00}; 
$\zeta_2$, pulsar: \cite{wil92}; 
$\zeta_3$, Solar-system: \cite{bv86,wil93}; 
$\eta$, $\Delta_{\rm net}$, Solar-system: \cite{dbf+94}, pulsar: \cite{wex00}; $
(\alpha_{c_1} - \alpha_0)^2$, pulsar: \cite{arz03}; 
$\dot G/G$, Solar-system: \cite{dbf+94,rea83,haa+83}, pulsar: \cite{tho96a}.
\label{table:PPN_params}}
\end{center}
\end{table}
}

\newpage
\subsection{Strong Equivalence Principle: Nordtvedt Effect}
\label{section:Delta}

The possibility of direct tests of the SEP through Lunar Laser Ranging
(LLR) experiments was first pointed out by Nordtvedt \cite{nor68b}.
As the masses of Earth and the Moon contain different fractional
contributions from self-gravitation, a violation of the SEP would
cause them to fall differently in the Sun's gravitational field.  This
would result in a ``polarization'' of the orbit in the direction of
the Sun.  LLR tests have set a limit of $|\eta| < 0.001$ (e.g.,
\cite{dbf+94,wil01}), where $\eta$ is a combination of PPN parameters:
\begin{equation}
\eta = 4\beta - \gamma -3 -\frac{10}{3}\xi-\alpha_1+\frac{2}{3}\alpha_2
-\frac{2}{3}\zeta_1 -\frac{1}{3}\zeta_2.
\label{equation:eta}
\end{equation}

The strong-field formalism instead uses the parameter $\Delta_{\rm i}$
\cite{ds91}, which for object ``{\it i}'' may be written as:
\begin{eqnarray}
\left(\frac{m_{\rm grav}}{m_{\rm inertial}}\right)_{i} & = & 1+\Delta_{\rm i} \nonumber \\
& = & 1+\eta\left(\frac{E^{\rm grav}}{mc^2}\right)_{i} + \eta^{\prime}\left(\frac{E^{\rm grav}}{mc^2}\right)_{i}^2 + \ldots.
\label{equation:Delta}
\end{eqnarray}
Pulsar--white dwarf systems then constrain $\Delta_{\rm net}=
\Delta_{\rm pulsar} - \Delta_{\rm companion}$ \cite{ds91}.  If the
SEP is violated, the equations of motion for such a system will
contain an extra acceleration $\Delta_{\rm net} {\bf g}$, where ${\bf
g}$ is the gravitational field of the Galaxy.  As the pulsar and the
white dwarf fall differently in this field, this $\Delta_{\rm net}
{\bf g}$ term will influence the evolution of the orbit of the system.
For low-eccentricity orbits, by far the largest effect will be a
long-term forcing of the eccentricity toward alignment with the
projection of ${\bf g}$ onto the orbital plane of the system. Thus the
time evolution of the eccentricity vector will not only depend on the
usual GR-predicted relativistic advance of periastron ($\dot \omega$)
but will also include a constant term.  Damour and Sch\"{a}fer
\cite{ds91} write the time-dependent eccentricity vector as:
\begin{equation}
{\bf e}(t) = {\bf e}_F +{\bf e}_R(t),
\end{equation}
where ${\bf e}_R(t)$ is the $\dot \omega$-induced rotating
eccentricity vector, and ${\bf e}_F$ is the forced component.  In
terms of $\Delta_{\rm net}$, the magnitude of ${\bf e}_F$ may be
written as \cite{ds91,wex97}:
\begin{equation}
|{\bf e}_F| = \frac{3}{2}\frac{\Delta_{\rm net} {\bf g}_{\perp}}{\dot \omega a (2\pi/P_{\rm b})},
\label{equation:ef1}
\end{equation}
where ${\bf g}_{\perp}$ is the projection of the gravitational field
onto the orbital plane, and $a = x/(\sin i)$ is the semi-major axis of
the orbit.  For small-eccentricity systems, this reduces to:
\begin{equation}
|{\bf e}_F| = \frac{1}{2}\frac{\Delta_{\rm net}{\bf g}_{\perp}c^2}{FG\,M (2\pi/P_{\rm b})^2},
\label{equation:ef2}
\end{equation}
where M is the total mass of the system, and, in GR, $F = 1$ and $G$
is Newton's constant.

Clearly, the primary criterion for selecting pulsars to test the SEP
is for the orbital system to have a large value of $P_{\rm b}^2/e$,
greater than or equal to $10^7$ days$^2$ \cite{wex97}.  However, as
pointed out by Damour and Sch{\"a}fer \cite{ds91} and Wex
\cite{wex97}, two age-related restrictions are also needed.  First of
all, the pulsar must be sufficiently old that the $\dot\omega$-induced
rotation of ${\bf e}$ has completed many turns and ${\bf e}_R(t)$ can
be assumed to be randomly oriented.  This requires that the
characteristic age $\tau_{\rm c} \gg 2\pi/\dot\omega$ and thus young
pulsars cannot be used.  Secondly, $\dot\omega$ itself must be larger
than the rate of Galactic rotation, so that the projection of ${\bf
g}$ onto the orbit can be assumed to be constant.  According to Wex
\cite{wex97}, this holds true for pulsars with orbital periods of less
than about 1000 days.

\begin{figure}[h]
  \def\epsfsize#1#2{0.4#1} 
\centerline{\epsfbox{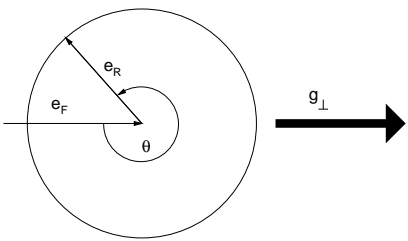}}
  \caption{\it ``Polarization'' of a nearly circular binary orbit
  under the influence of a forcing vector ${\bf g}$, showing the
  relation between the forced eccentricity, ${\bf e}_F$, the
  eccentricity evolving under the general-relativistic advance of
  periastron, ${\bf e}_R(t)$, and the angle $\theta$.  After \cite{wex97}.}
  \label{figure:eccrotate}
\end{figure}

Converting equation~\ref{equation:ef2} to a limit on $\Delta_{\rm
net}$ requires some statistical arguments to deal with the unknowns in
the problem.  First is the actual component of the observed
eccentricity vector (or upper limit) along a given direction.  Damour
and Sch\"{a}fer (1991)
\cite{ds91} assume the worst case of possible cancellation between the
two components of ${\bf e}$, namely that $|{\bf e}_F| \simeq |{\bf
e}_R|$.  With an angle $\theta$ between ${\bf g}_{\perp}$ and ${\bf
e}_R$ (see Figure~\ref{figure:eccrotate}), they write $|{\bf e}_F|
\leq e/(2\sin(\theta/2))$.  Wex \cite{wex97,wex00} corrects this
slightly and uses the inequality:
\begin{equation}
|{\bf e}_F| \leq e \xi_1(\theta), \,\,\,\,\,
\xi_1(\theta) = \left [ \begin{array}{r@{\quad:\quad}l}
        1/\sin\theta & \theta \in [0,\pi/2) \\
	1 & \theta \in [\pi/2,3\pi/2] \\
	-1/\sin\theta & \theta \in (3\pi/2,2\pi)
	\end{array}  \right. ,
\label{equation:xi1}
\end{equation}
where $e = |\bf{e}|$.  In both cases $\theta$ is assumed to have a
uniform probability distribution between 0 and $2\pi$.

Next comes the task of estimating the projection of ${\bf g}$ onto the
orbital plane.  The projection can be written as:
\begin{equation}
|{\bf g}_{\perp}| = |{\bf g}|[1-(\cos i \cos\lambda + \sin i \sin\lambda\sin\Omega)^2]^{1/2},
\label{equation:gperp}
\end{equation}
where $i$ in the inclination angle of the orbital plane relative to
the line of sight, $\Omega$ is the line of nodes and $\lambda$ is the angle
between the line of sight to the pulsar and ${\bf g}$ \cite{ds91}.
The values of $\lambda$ and $|{\bf g}|$ can be determined from models
of the Galactic potential (e.g., \cite{kg89,am86}).  The inclination
angle $i$ can be estimated if even crude estimates of the neutron star
and companion masses are available, from statistics of NS masses
(e.g., \cite{tc99}) and/or a relation between the size of the orbit
and the WD companion mass (e.g.,\cite{rpj+95}).  However, the angle
$\Omega$ is also usually unknown and must also be assumed to be
uniformly distributed between 0 and $2\pi$.

Damour and Sch\"{a}fer \cite{ds91} use the PSR~B1953+29 system and
integrate over the angles $\theta$ and $\Omega$ to determine a 90\%
confidence upper limit of $\Delta_{\rm net} < 1.1\times10^{-2}$.  Wex
\cite{wex97} uses an ensemble of pulsars, calculating for each system
the probability (fractional area in $\theta$--$\Omega$ space) that
$\Delta_{\rm net}$ is less than a given value, and then deriving a
cumulative probability for each value of $\Delta_{\rm net}$.  In this
way he derives $\Delta_{\rm net} < 5\times10^{-3}$ at 95\% confidence.
However, this method may be vulnerable to selection effects: perhaps
the observed systems are not representative of the true population.
Wex \cite{wex00} later overcomes this problem by inverting the
question.  Given a value of $\Delta_{\rm net}$, an upper limit on
$|\theta|$ is obtained from equation~\ref{equation:xi1}.  A Monte
Carlo simulation of the expected pulsar population (assuming a range
of masses based on evolutionary models and a random orientation of
$\Omega$) then yields a certain fraction of the population that agree
with this limit on $|\theta|$.  The collection of pulsars ultimately
gives a limit of $\Delta_{\rm net} < 9\times10^{-3}$ at 95\%
confidence.  This is slightly weaker than Wex's previous limit but
derived in a more rigorous manner.

Prospects for improving the limits come from the discovery of new
suitable pulsars, and from better limits on eccentricity from
long-term timing of the current set of pulsars.  In principle,
measurement of the full orbital orientation (i.e., $\Omega$ and $i$)
for certain systems could reduce the dependence on statistical
arguments.  However, the possibility of cancellation between $|{\bf
e}_F|$ and $|{\bf e}_R|$ will always remain.  Thus, even though the
required angles have in fact been measured for the millisecond pulsar
J0437$-$4715 \cite{vbb+01}, its comparatively large observed
eccentricity of $\sim 2\times 10^{-5}$ means it will not significantly
affect the current limits.
 
\newpage
\subsection{Preferred-Frame Effects and Non-conservation of Momentum}
\label{section:frames}

\subsubsection{Limits on $\hat {\alpha_1}$}
\label{section:alpha1}

A non-zero $\hat {\alpha_1}$ implies that the velocity ${\bf w}$ of a
binary pulsar system (relative to a ``universal'' background reference
frame given by the Cosmic Microwave Background, or CMB) will affect
its orbital evolution.  In a manner similar to the effects of a
non-zero $\Delta_{\rm net}$, the time evolution of the eccentricity
will depend on both $\dot \omega$ and a term which tries to force the
semi-major axis of the orbit to align with the projection of the
system velocity onto the orbital plane.

The analysis proceeds in a similar fashion to that for $\Delta_{\rm
net}$, except that the magnitude of ${\bf e}_F$ is now written as
\cite{de92a,bcd96}:
\begin{equation}
|{\bf e}_F| = \frac{1}{12}\hat {\alpha_1}\left |\frac{m_1-m_2}{m_1+m_2}\right|
\frac{|w_{\perp}|}{\left[G(m_1+m_2)(2\pi/P_{\rm b})\right]^{1/3}},
\label{equation:efalpha1}
\end{equation}
where $w_{\perp}$ is the projection of the system velocity onto the
orbital plane.  The angle $\lambda$, used in determining this
projection in a manner similar to that of
equation~\ref{equation:gperp}, is now the angle between the line of
sight to the pulsar and the absolute velocity of the binary system.

The figure of merit for systems used to test $\hat {\alpha_1}$ is
$P_{\rm b}^{1/3}/e$. As for the $\Delta_{\rm net}$ test, the systems
must be old, so that $\tau_{\rm c} \gg 2\pi/\dot\omega$, and
$\dot\omega$ must be larger than the rate of Galactic rotation.
Examples of suitable systems are PSR~J2317+1439 \cite{cnt93, bcd96}
with a last published value of $e < 1.2\times10^{-6}$ in 1996
\cite{cnt96}, and PSR~J1012+5307, with $e < 8\times10^{-7}$
\cite{lcw+01}.  This latter system is especially valuable because
observations of its white-dwarf component yield a radial velocity
measurement \cite{cgk98}, eliminating the need to find a lower limit
on an unknown quantity.  The analysis of Wex \cite{wex00} yields a
limit of $\hat {\alpha_1} < 1.4 \times 10^{-4}$.  This is comparable
in magnitude to the weak-field results from lunar laser ranging, but
incorporates strong field effects as well.

\newpage 
\subsubsection{Limits on $\hat \alpha_3$}
\label{section:alpha3}

Tests of $\hat \alpha_3$ can be derived from both binary and single
pulsars, using slightly different techniques.  A non-zero $\hat
\alpha_3$, which implies both a violation of local Lorentz invariance
and non-conservation of momentum, will cause a rotating body to
experience a self-acceleration ${\bf a}_{\rm self}$ in a direction
orthogonal to both its spin ${\bf \Omega}_{\rm S}$ and its absolute
velocity ${\bf w}$ \cite{nw72}:
\begin{equation}
{\bf a}_{\rm self} = -\frac{1}{3}\hat \alpha_3 \frac{E^{\rm grav}}{(m\,c^2)}{\bf w}
\times {\bf \Omega}_{\rm S}.
\label{alpha3accel}
\end{equation}
Thus the self-acceleration depends strongly on the compactness of the
object, as discussed in Section~\ref{section:sep} above.

An ensemble of single (isolated) pulsars can be used to set a limit on
$\hat \alpha_3$ in the following manner.  For any given pulsar, it is
likely that some fraction of the self-acceleration will be directed
along the line of sight to the Earth.  Such an acceleration will
contribute to the observed period derivative $\dot P$ via the Doppler
effect, by an amount:
\begin{equation}
\dot P_{\hat \alpha_3} = \frac{P}{c} {\bf \hat n}\cdot{\bf a}_{\rm self},
\label{equation:pdotalpha3}
\end{equation}
where ${\bf \hat n}$ is a unit vector in the direction from the pulsar
to the Earth.  The analysis of Will \cite{wil93} assumes random
orientations of both the pulsar spin axes and velocities, and finds
that, on average, $|\dot P_{\hat \alpha_3}| \simeq 5 \times
10^{-5}|\hat \alpha_3|$, independent of the pulse period.  The {\it
sign} of the $\hat \alpha_3$ contribution to $\dot P$, however, may be
positive or negative for any individual pulsar, thus if there were a
large contribution on average, one would expect to observe pulsars
with both positive and negative total period derivatives.  Young
pulsars in the field of the Galaxy (pulsars in globular clusters
suffer from unknown accelerations from the cluster gravitational
potential and do not count toward this analysis) all show positive
period derivatives, typically around $10^{-14}$s/s.  Thus the maximum
possible contribution from $\hat \alpha_3$ must also be considered to
be of this size, and the limit is given by $|\hat \alpha_3| <
2\times10^{-10}$ \cite{wil93}.

Bell \cite{bel96} applies this test to a set of millisecond
pulsars; these have much smaller period derivatives, on the order of
$10^{-20}$s/s.  Here it is also necessary to account for the
``Shklovskii effect'' \cite{shk70} in which a similar Doppler-shift
addition to the period derivative results from the transverse motion
of the pulsar on the sky: 
\begin{equation}
\dot P_{\rm pm} = P\mu^2\frac{d}{c}
\label{equation:pdotpm}
\end{equation}
where $\mu$ is the proper motion of the pulsar and $d$ is the distance
between the Earth and the pulsar.  The distance is usually poorly
determined, with uncertainties of typically 30\% resulting from models
of the dispersive free electron density in the Galaxy
\cite{tc93,cl02}.  Nevertheless, once this correction (which is
always positive) is applied to the observed period derivatives for
isolated millisecond pulsars, a limit on $|\hat \alpha_3|$ on the order of
$10^{-15}$ results \cite{bel96,bd96}.

In the case of a binary pulsar--white-dwarf system, both bodies
experience a self-acceleration.  The combined accelerations affect
both the velocity of the centre of mass of the system (an effect which
may not be readily observable) and the relative motion of the two
bodies \cite{bd96}.  The relative-motion effects break down into a
term involving the coupling of the spins to the absolute motion of the
centre of mass, and a second term which couples the spins to the
orbital velocities of the stars.  The second term induces only a very
small, unobservable correction to $P_{\rm b}$ and $\dot \omega$
\cite{bd96}.  The first term, however, can lead to a significant
test of $\hat \alpha_3$.  Both the compactness and the spin of the
pulsar will completely dominate those of the white dwarf, making the
net acceleration of the two bodies effectively:
\begin{equation}
{\bf a}_{\rm self} = \frac{1}{6}\hat \alpha_3 c_{\rm p}{\bf w}
\times {\bf \Omega}_{\rm Sp},
\label{alpha3accel2}
\end{equation}
where $c_{\rm p}$ and ${\bf \Omega}_{\rm Sp}$ denote the compactness
and spin angular frequency of the pulsar, respectively, and ${\bf w}$
is the velocity of the system.  For evolutionary reasons (e.g.,
\cite{bv91}), the spin axis of the pulsar may be assumed to be aligned
with the orbital angular momentum of the system, hence the net effect
of the acceleration will be to induce a polarization of the
eccentricity vector within the orbital plane.  The forced eccentricity
term may be written as:
\begin{equation}
|{\bf e}_F| = \hat \alpha_3\frac{c_{\rm p}|{\bf w}|}{24\pi}\frac{P_{\rm b}^2}{P}\frac{c^2}{G(m_1+m_2)}\sin\beta
\label{equation:efalpha3}
\end{equation}
where $\beta$ is the (unknown) angle between ${\bf w}$ and ${\bf
\Omega}_{\rm Sp}$, and $P$ is, as usual, the spin period of the
pulsar: $P=2\pi/\Omega_{\rm Sp}$.

The figure of merit for systems used to test $\hat \alpha_3$ is $P_{\rm
b}^2/(eP)$.  The additional requirements of $\tau_{\rm c} \gg
2\pi/\dot\omega$ and $\dot\omega$ being larger than the rate of
Galactic rotation also hold.  The 95\% confidence limit derived by Wex
\cite{wex00} for an ensemble of binary pulsars is $\hat \alpha_3 <
1.5\times 10^{-19}$, much more stringent than for the single-pulsar
case.

\newpage
\subsubsection{Limits on $\zeta _2$}
\label{section:zeta2}

Another PPN parameter that predicts the non-conservation of momentum
is $\zeta _2$.  It will contribute, along with $\alpha_3$, to an
acceleration of the centre of mass of a binary system
\cite{wil92,wil93}:
\begin{equation}
{\bf a}_{\rm CM} = (\alpha_3+\zeta_2) \frac{\pi m_1m_2(m_1-m_2)}{P_b[(m_1+m_2)a(1-e^2)]^{3/2}} e\, {\bf n}_{\rm p},
\label{equation:zeta2}
\end{equation}
where ${\bf n}_{\rm p}$ is a unit vector from the centre of mass to the
periastron of $m_1$.  This acceleration produces the same type of
Doppler-effect contribution to a binary pulsar's $\dot P$ as described
in Section~\ref{section:alpha3}.  In a small-eccentricity
system, this contribution would not be separable from the $\dot P$
intrinsic to the pulsar.  However, in a highly eccentric binary such
as PSR~B1913+16, the longitude of periastron advances significantly ---
for PSR~B1913+16, it has advanced nearly 120$^{\circ}$ since the
pulsar's discovery.  In this case, the projection of ${\bf a}_{\rm
CM}$ along the line of sight to the Earth will change considerably
over the long term, producing an effective {\it second} derivative of
the pulse period.  This $\ddot P$ is given by \cite{wil92,wil93}:
\begin{equation}
\ddot P = \frac{P}{2}(\alpha_3+\zeta_2)m_2\sin i \left(\frac{2\pi}{P_{\rm b}}\right)^2
\frac{X(1-X)}{(1+X)^2} \frac{e\, \dot\omega\, \cos \omega}{(1-e^2)^{3/2}},
\label{equation:pddotfromzeta2}
\end{equation}
where $X = m_1/m_2$ is the mass ratio of the two stars and an average
value of $\cos \omega$ is chosen.  As of 1992, the 95\% confidence
upper limit on $\ddot P$ was $4\times 10^{-30}$s$^{-1}$
\cite{tw89,wil92}.  This leads to an upper limit on
$(\alpha_3+\zeta_2)$ of $4\times 10^{-5}$ \cite{wil92}.  As $\alpha_3$
is orders of magnitude smaller than this (see
Section~\ref{section:alpha3}), this can be interpreted as a limit on
$\zeta_2$ alone.  Although PSR~B1913+16 is of course still observed,
the infrequent campaign nature of the observations makes it difficult
to set a much better limit on $\ddot P$ (J. Taylor, private
communication, as cited in \cite{kws03}).  The other well-studied
double-neutron-star binary, PSR~B1534+12, yields a weaker test due to
its orbital parameters and very similar component masses.  A
complication for this test is that an observed $\ddot P$ could also be
interpreted as timing noise (sometimes seen in recycled pulsars
\cite{ktr94}) or else a manifestation of profile changes due to
geodetic precession \cite{kdg96,kws03}.

\newpage
\subsection{Strong Equivalence Principle: Dipolar Gravitational Radiation}
\label{section:dipolar}

General Relativity predicts gravitational radiation from the
time-varying mass quadrupole of a binary pulsar system.  The
spectacular confirmation of this prediction will be discussed in
section~\ref{section:strong_field} below.  GR does not, however,
predict {\it dipolar} gravitational radiation, though many theories
that violate the SEP do.  In these theories, dipolar gravitational
radiation results from the difference in gravitational binding energy
of the two components of a binary.  For this reason, neutron
star--white dwarf binaries are the ideal laboratories to test the
strength of such dipolar emission.  The expected rate of change of the
period of a circular orbit due to dipolar emission can be written as
\cite{wil93,de96b}:
\begin{equation}
\dot P_{\rm b\,Dipole} =-\frac{4\pi^2G_{\ast}}{c^3\,P_{\rm b}}\frac{m_1m_2}{m_1+m_2}(\alpha_{c_1}-\alpha_{c_2})^2,
\label{equation:pbdotdipole}
\end{equation}
where $G_{\ast} = G$ in GR, and $\alpha_{c_{i}}$ is the coupling
strength of body ``{\it i}'' to a scalar gravitational field
\cite{de96b}.  (Similar expressions can be derived when casting
$\dot P_{\rm b\,Dipole}$ in terms of the parameters of specific
tensor-scalar theories, such as Brans-Dicke theory
\cite{bd61}. Equation~\ref{equation:pbdotdipole}, however, tests a
more general class of theories.)  The best test systems here are of
course pulsar--white dwarf binaries with short orbital periods, such
as PSR~B0655+64 and PSR~J1012+5307, where
$\alpha_{c_1}\gg\alpha_{c_2}$ so that a strong limit can be set on the
coupling of the pulsar itself.  For PSR~B0655+64, Damour and
Esposito-Far\`{e}se \cite{de96b} used the observed limit of $\dot
P_{\rm b} = (1\pm4)\times 10^{-13}$ \cite{arz95} to derive
$(\alpha_{c_1} -\alpha_0)^2 < 3\times10^{-4}$ (1-$\sigma)$, where
$\alpha_0$ is a reference value of the coupling at infinity.  More
recently, Arzoumanian \cite{arz03} has set a 2-$\sigma$ upper limit of
$|\dot P_{\rm b}/P_{\rm b}| < 1\times 10^{-10}$\,yr$^{-1}$, or $|\dot
P_{\rm b}| < 2.7\times 10^{-13}$, which yields $(\alpha_{c_1}
-\alpha_0)^2 < 2.7\times10^{-4}$, somewhat tighter.  For
PSR~J1012+5307, a ``Shklovskii'' correction (see \cite{shk70} and
Section~\ref{section:alpha3}) for the transverse motion of the system
and a correction for the (small) predicted amount of quadrupolar
radiation must first be subtracted from the observed upper limit to
arrive at $\dot P_{\rm b} = (-0.6\pm1.1)\times 10^{-13}$ and
$(\alpha_{c_1} -\alpha_0)^2 < 4\times10^{-4}$ at 95\% confidence
\cite{lcw+01}.  It should be noted that both these limits depend on
estimates of the masses of the two stars and do not address the
(unknown) equation of state of the neutron stars.

Limits may also be derived from double-neutron-star systems (e.g.,
\cite{wil77,wz89}), although here the difference in the coupling
constants is small and so the expected amount of dipolar radiation is
also small compared to the quadrupole emission.  However, certain
alternative gravitational theories in which the quadrupolar radiation
predicts a {\it positive} orbital period derivative independently of
the strength of the dipolar term (e.g., \cite{ros73,ni73,ll73}) are
ruled out by the observed decreasing orbital period in these systems
\cite{wt81}.

Other pulsar--white dwarf systems with short orbital periods are
mostly found in globular clusters, where the cluster potential will
also contribute to the observed $\dot P_{\rm b}$, or in interacting
systems, where tidal effects or magnetic braking may affect the
orbital evolution (e.g., \cite{as94,es00,nat00}).  However, one system
that offers interesting prospects is the recently discovered
PSR~J1141$-$6545 \cite{klm+00a} which is a young pulsar with
white-dwarf companion in a 4.75-hour orbit.  In this case, though, the
pulsar was formed {\it after} the white dwarf, instead of being
recycled by the white-dwarf progenitor, and so the orbit is still
highly eccentric.  This system is therefore expected both to emit
sizable amounts of quadrupolar radiation --- $\dot P_{\rm b}$ could
be measurable as soon as 2004 \cite{klm+00a} --- and to be a good test
candidate for dipolar emission \cite{gw02}.

\newpage
\subsection{Preferred-location Effects: Variation of Newton's Constant}
\label{section:Gdot}

Theories that violate the SEP by allowing for preferred locations (in
time as well as space) may permit Newton's constant, $G$, to vary.  In
general, variations in $G$ are expected to occur on the timescale of
the age of the Universe, such that $\dot G/G \sim H_0 \sim 0.7\times
10^{-10}$\,yr$^{-1}$, where $H_0$ is the Hubble constant.  Three
different pulsar-derived tests can be applied to these predictions, as
a SEP-violating time-variable $G$ would be expected to alter the
properties of neutron stars and white dwarfs, and to affect binary
orbits.

\newpage
\subsubsection{Spin Tests}
\label{Gdotspin}

By affecting the gravitational binding of neutron stars, a non-zero
$\dot G$ would reasonably be expected to alter the moment of inertia
of the star and hence change its spin on the same timescale
\cite{cs68}.  Goldman \cite{gol90} writes:
\begin{equation}
\left(\frac{\dot P}{P}\right)_{\dot G} = \left(\frac{\partial \ln I}{\partial \ln G}\right)_N\frac{\dot G}{G},
\label{PdotfromGdot}
\end{equation}
where I is the moment of inertia of the neutron star, about
$10^{45}$\,g\,cm$^2$, and $N$ is the (conserved) total number of
baryons in the star.  By assuming that this represents the {\it only}
contribution to the observed $\dot P$ of PSR~B0655+64, in a manner
reminiscent of the test of $\hat \alpha_3$ described above, Goldman then
derives an upper limit of $|\dot G/G| \le (2.2-5.5)\times
10^{-11}$\,yr$^{-1}$, depending on the stiffness of the neutron star
equation of state.  Arzoumanian \cite{arz95} applies similar
reasoning to PSR~J2019+2425 \cite{ntf93} which has a characteristic
age of 27 Gyr once the ``Shklovskii'' correction is applied \cite{nt95}.
Again depending on the equation of state, the upper limits from this
pulsar are $|\dot G/G| \le (1.4-3.2)\times 10^{-11}$\,yr$^{-1}$
\cite{arz95}.  These values are similar to those obtained by
solar-system experiments such as laser ranging to the Viking Lander on
Mars (e.g., \cite{rea83,haa+83}).  Several other millisecond pulsars,
once ``Shklovskii'' and Galactic-acceleration corrections are taken
into account, have similarly large characteristic ages
(e.g. \cite{cnt96,tsb+99}).  

\newpage
\subsubsection{Orbital Decay Tests}
\label{section:Gdotorb}

The effects on the orbital period of a binary system of a varying $G$
were first considered by Damour, Gibbons \& Taylor \cite{dgt88}, who
expected:
\begin{equation}
\left(\frac{\dot P_{\rm b}}{P_{\rm b}}\right)_{\dot G} = -2\frac{\dot G}{G}.
\label{equation:PbdotfromGdot1}
\end{equation}
Applying this equation to the limit on the deviation from GR of the
$\dot P_{\rm b}$ for PSR~1913+16, they found a value of $\dot G/G =
(1.0 \pm 2.3)\times 10^{-11}$\,yr$^{-1}$.  Nordtvedt 
\cite{nor90} took into account the effects of $\dot G$ on
neutron-star structure, realizing that the total mass and angular
momentum of the binary system would also change.  The corrected
expression for $\dot P_{\rm b}$ incorporates the compactness parameter
$c_{\rm i}$ and is:
\begin{equation}
\left(\frac{\dot P_{\rm b}}{P_{\rm b}}\right)_{\dot G} = -\left[2-\left(
\frac{m_1c_1+m_2c_2}{m_1+m_2}\right) -\frac{3}{2}\left(\frac{m_1c_2+m_2c_1}{m_1+m_2}
\right)\right]\frac{\dot G}{G}.
\label{equation:PbdotfromGdot2}
\end{equation}
(Note that there is a difference of a factor of $-2$ in Nordtvedt's
definition of $c_{i}$ versus the Damour definition used throughout
this article.)  Nordtvedt's corrected limit for PSR~B1913+16 is
therefore slightly weaker.  A better limit actually comes from the
neutron-star--white dwarf system PSR~B1855+09, with a measured limit
on $\dot P_{\rm b}$ of $(0.6 \pm 1.2)\times10^{-12}$ \cite{ktr94}.
Using equation~\ref{equation:PbdotfromGdot1} this leads to a bound of
$\dot G/G = (-9 \pm 18)\times 10^{-12}$\,yr$^{-1}$, which Arzoumanian
\cite{arz95} corrects using equation~\ref{equation:PbdotfromGdot2}
and an estimate of NS compactness to $\dot G/G = (-1.3 \pm 2.7)\times
10^{-11}$\,yr$^{-1}$.  Prospects for improvement come directly from
improvements to the limit on $\dot P_{\rm b}$.  Even though
PSR~J1012+5307 has a tighter limit on $\dot P_{\rm b}$
\cite{lcw+01}, its shorter orbital period means that the $\dot G$
limit it sets is a factor of 2 weaker than for PSR~B1855+09.

\newpage
\subsubsection{Changes in the Chandrasekhar Mass}
\label{Gdotchandra}

\begin{figure}[h]
  \def\epsfsize#1#2{0.4#1} 
\centerline{\epsfbox{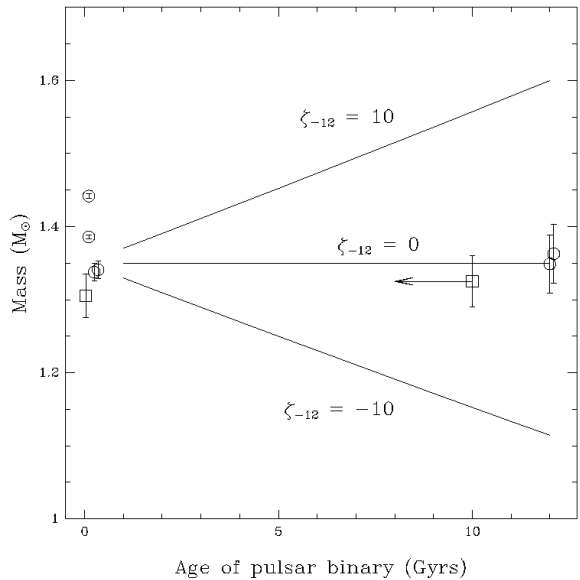}}
  \caption{\it Measured neutron star masses as a function of age.
  The solid lines show predicted changes in the average neutron star
  mass corresponding to hypothetical variations in $G$, where
  $\zeta_{-12} = 10$ implies $\dot G/G =10\times10^{-12}$\,yr$^{-1}$.
  From \cite{tho96a}, used by permission.}\label{figure:Gdotchandra}
\end{figure}

The Chandrasekhar mass, $M_{\rm CH}$, is the maximum mass possible for
a white dwarf supported against gravitational collapse by electron
degeneracy pressure \cite{cha31}. Its value --- about $1.4\,
M_{\odot}$ --- comes directly from Newton's constant: $M_{\rm CH} \sim
(\hbar\,c/G)^{3/2}/m_{\rm N}^2$, where $\hbar$ is Planck's constant
and $m_{\rm N}$ is the neutron mass.  All measured and constrained
pulsar masses are consistent with a narrow distribution centred very
close to $M_{\rm CH}$: $1.35\pm0.04\,M_{\odot}$ \cite{tc99}.  Thus
it is reasonable to assume that $M_{\rm CH}$ sets the typical neutron
star mass, and to check for any changes in the average neutron star
mass over the lifetime of the Universe.  Thorsett \cite{tho96a}
compiles a list of measured and average masses from 5
double-neutron-star binaries with ages ranging from 0.1 Gyr to 12 or
13 Gyr in the case of the globular-cluster binary B2127+11C.  Using a
Bayesian analysis, he finds a limit of $\dot G/G = (-0.6\pm4.2)\times
10^{-12}$\,yr$^{-1}$ at the 95\% confidence level, the strongest limit
on record.  Figure~\ref{figure:Gdotchandra} illustrates the logic
applied.

While some cancellation of ``observed'' mass changes might be expected
from the changes in neutron-star binding energy
(cf. Section~\ref{section:Gdotorb} above), these will be smaller than
the $M_{\rm CH}$ changes by a factor of order the compactness and can
be neglected.  Also, the claimed variations of the fine structure
constant of order $\Delta \alpha/\alpha \simeq
-0.72\pm0.18\times10^{-5}$ \cite{wmf+01} over the redshift range $0.5
< z < 3.5$ could introduce a maximum derivative of $d(\hbar
c)/dt/(\hbar c)$ of about $5\times10^{-16}$\,yr$^{-1}$ and hence
cannot influence the Chandrasekhar mass at the same level as the
hypothesized changes in $G$.

One of the 5 systems used by Thorsett has since been shown to have a
white-dwarf companion \cite{vk99}, but as this is one of the
youngest systems, this will not change the results appreciably.  The
recently discovered PSR~J1811$-$1736 \cite{lcm+00}, a
double-neutron-star binary, has characteristic age only $\tau_{\rm c}
\sim 1\,$Gyr and therefore will also not significantly strengthen the
limit.  Ongoing searches for pulsars in globular clusters stand the
best chance of discovering old double-neutron-star binaries for which
the component masses can eventually be measured.

\newpage
\section{Tests of GR --- Strong-Field Gravity}
\label{section:strong_field}

The best-known uses of pulsars for testing the predictions of
gravitational theories are those in which the predicted strong-field
effects are compared directly against observations.  As essentially
point-like objects in strong gravitational fields, neutron stars in
binary systems provide extraordinarily clean tests of these
predictions.  This section will cover the relation between the
``post-Keplerian'' timing parameters and strong-field effects, and
then discuss the three binary systems which yield complementary
high-precision tests.

\newpage
\subsection{Post-Keplerian Timing Parameters}
\label{section:pkparms}

In any given theory of gravity, the post-Keplerian (PK) parameters can
be written as functions of the pulsar and companion star masses and
the Keplerian parameters.  As the two stellar masses are the only
unknowns in the description of the orbit, it follows that measurement
of any two PK parameters will yield the two masses, and that
measurement of three or more PK parameters will over-determine the
problem and allow for self-consistency checks.  It is this test for
internal consistency among the PK parameters that forms the basis of
the classic tests of strong-field gravity.  It should be noted that
the basic Keplerian orbital parameters are well-measured and can
effectively be treated as constants here.

In general relativity, the equations describing the PK parameters in
terms of the stellar masses are (see \cite{dd86,tw89,dt92}):
\begin{equation}
\dot\omega = 3 \left(\frac{P_b}{2\pi}\right)^{-5/3}
  (T_\odot M)^{2/3}\,(1-e^2)^{-1}\,, \label{eq:omdot} 
\end{equation}
\begin{equation}
\gamma = e \left(\frac{P_b}{2\pi}\right)^{1/3}
  T_\odot^{2/3}\,M^{-4/3}\,m_2\,(m_1+2m_2) \,, 
\end{equation}
\begin{equation}
\dot P_b = -\,\frac{192\pi}{5}
  \left(\frac{P_b}{2\pi}\right)^{-5/3}
  \left(1 + \frac{73}{24} e^2 + \frac{37}{96} e^4 \right)
  (1-e^2)^{-7/2}\,T_\odot^{5/3}\, m_1\, m_2\, M^{-1/3}\,,
  \label{eq:pbdot} 
\end{equation}
\begin{equation}
r = T_\odot\, m_2\,, 
\end{equation}
\begin{equation}
s = x \left(\frac{P_b}{2\pi}\right)^{-2/3}
  T_\odot^{-1/3}\,M^{2/3}\,m_2^{-1}\,. \label{eq:s}
\end{equation}
where $s\equiv\sin i$, $M = m_1+m_2$ and $T_\odot\equiv GM_\odot/c^3 =
4.925490947\,\mu$s.  Other theories of gravity, such as those with one
or more scalar parameters in addition to a tensor component, will have
somewhat different mass dependencies for these parameters.  Some
specific examples will be discussed in Section~\ref{section:combined}
below.

\newpage
\subsection{The Original System: PSR B1913+16}
\label{section:psr1913}

The prototypical double-neutron-star binary PSR~B1913+16 was
discovered at the Arecibo Observatory \cite{arecibo} in 1974
\cite{ht75a}.  Over nearly 30 years of timing, its system parameters
have shown a remarkable agreement with the predictions of GR, and in
1993 Hulse and Taylor received the Nobel Prize in Physics for its
discovery \cite{hul94,tay94b}.  In the highly eccentric 7.75-hour
orbit, the two neutron stars are separated by only 3.3 light-seconds
and have velocities up to 400 km/s.  This provides an ideal laboratory
for investigating strong-field gravity.

\begin{table}[h]
\begin{center}
\begin{tabular} {l l}
\hline 
Parameter & Value \\
\hline\hline
Orbital period, $P_b$ (d)   & 0.322997462727(5)  \\ 
Projected semi-major axis, $x$ (s)  & 2.341774(1) \\
Eccentricity, $e$   & 0.6171338(4) \\ 
Longitude of periastron, $\omega$ (deg)  & 226.57518(4)  \\
Epoch of periastron, $T_0$ (MJD)  &  46443.99588317(3)\\
 & \\
Advance of periastron, $\dot\omega$ (deg\,yr$^{-1}$)
  & 4.226607(7) \\
Gravitational redshift, $\gamma$ (ms)   & 4.294(1)  \\
Orbital period derivative, $(\dot P_b)^{\rm obs}$ $(10^{-12})$   &
  $-$2.4211(14) \\
\hline
\end{tabular}
\caption{Orbital parameters for PSR~B1913+16 in the DD framework, taken from \cite{wt03}.\label{table:params_1913}
}
\end{center}
\end{table}

\begin{figure}[h]
  \def\epsfsize#1#2{0.4#1} 
\centerline{\epsfbox{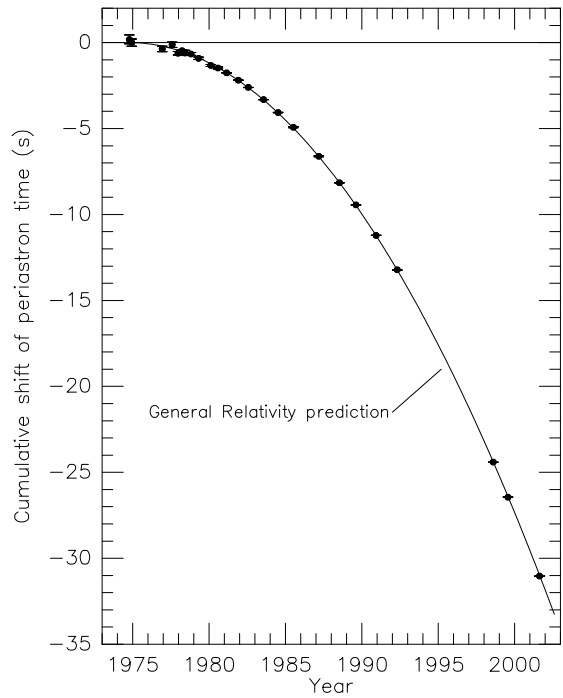}}
  \caption{\it The parabola indicates the predicted accumulated shift
  in the time of periastron for PSR~B1913+16, caused by the decay of
  the orbit.  The measured values of the epoch of periastron are
  indicated by the data points.  From \cite{wt03}, courtesy Joel
  Weisberg.}\label{figure:t0_1913}
\end{figure}

For PSR~B1913+16, three PK parameters are well measured: the combined
gravitational redshift and time dilation parameter $\gamma$, the
advance of periastron $\dot \omega$ and the derivative of the orbital
period, $\dot P_{\rm b}$.  The orbital parameters for this pulsar,
measured in the theory-independent ``DD'' system, are listed in
Table~\ref{table:params_1913} \cite{tw89,wt03}.

The task is now to judge the agreement of these parameters with GR.  A
second useful timing formalism is ``DDGR'' \cite{dd86,dt92}, which
assumes GR to be the true theory of gravity and fits for the total and
companion masses in the system, using these quantities to calculate
``theoretical'' values of the PK parameters.  Thus one can make a
direct comparison between the measured DD PK parameters and the values
predicted by DDGR using the same data set; the parameters for
PSR~B1913+16 agree with their predicted values to better than 0.5\%
\cite{tw89}.  The classic demonstration of this agreement is shown
in Figure~\ref{figure:t0_1913} \cite{wt03}, in which the observed
accumulated shift of periastron is compared to the predicted amount.

\begin{figure}[h]
  \def\epsfsize#1#2{0.4#1} 
\centerline{\epsfbox{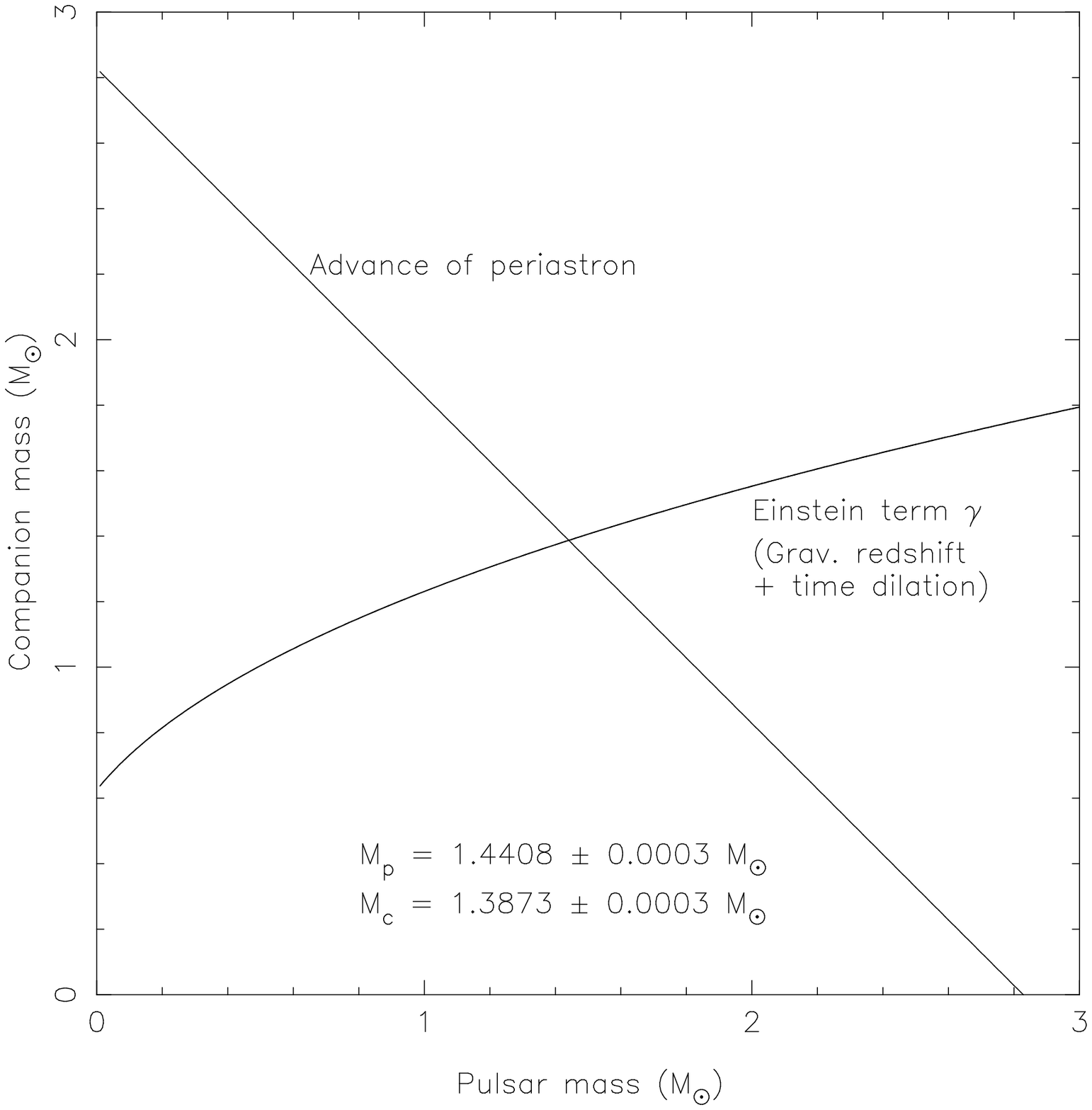}}
  \caption{\it Mass-mass diagram for the PSR B1913+16 system, using
  the $\dot \omega$ and $\gamma$ parameters listed in
  Table~\ref{table:params_1913}.  The uncertainties are smaller than
  the widths of the lines.  The lines intersect at the point given by
  the masses derived under the DDGR formalism.  From \cite{wt03},
  courtesy Joel Weisberg.}\label{figure:m1m2_1913}
\end{figure}

In order to check the self-consistency of the overdetermined set of
equations relating the PK parameters to the neutron star masses, it is
helpful to plot the allowed $m_1-m_2$ curves for each parameter and to
verify that they intersect at a common point.
Figure~\ref{figure:m1m2_1913} displays the $\dot \omega$ and $\gamma$
curves for PSR~B1913+16; it is clear that the curves do intersect, at
the point derived from the DDGR mass predictions.

Clearly any theory of gravity which does not pass such a
self-consistency test can be ruled out.  However, it is possible to
construct alternate theories of gravity which, while producing very
different curves in the $m_1-m_2$ plane, do pass the PSR~B1913+16 test
and possibly weak-field tests as well \cite{de92}.  Such theories
are best dealt with by combining data from multiple pulsars as well as
solar-system experiments (see Section~\ref{section:combined}).

A couple of practical points are worth mentioning.  The first is that
the unknown radial velocity of the binary system relative to the Solar
System Barycentre (SSB) will necessarily induce a Doppler shift in the
orbital and neutron-star spin periods.  This will change the observed
stellar masses by a small fraction but will cancel out of the
calculations of the PK parameters \cite{dd86}.  The second is that
the measured value of the orbital period derivative, $\dot P_{\rm b}$,
{\it is} contaminated by several external contributions.  Damour and
Taylor \cite{dt91} consider the full range of possible contributions
to $\dot P_{\rm b}$ and calculate values for the two most important:
the acceleration of the pulsar binary centre of mass relative to the
SSB in the Galactic potential, and the ``Shklovskii'' $v^2/r$ effect
due to the transverse proper motion of the pulsar
(cf. Section~\ref{section:alpha3}).  Both of these contributions have
been subtracted from the measured value of $\dot P_{\rm b}$ before it
is compared with the GR prediction.  It is our current knowledge of
the Galactic potential and the resulting models of Galactic
acceleration (e.g., \cite{kg89,am86}) which now limit the precision of
the test of GR resulting from this system.

\newpage
\subsection{PSR B1534+12 and Other Binary Pulsars}
\label{section:psr1534}

A second double-neutron-star binary, PSR~B1534+12, was discovered
during a drift-scan survey at Arecibo Observatory in 1990
\cite{wol91a}.  This system is quite similar to PSR~B1913+16: it
also has a short (10.1-hour) orbit, though it is slightly wider and
less eccentric.  PSR~B1534+12 does possess some notable advantages
relative to its more famous cousin: it is closer to the Earth and
therefore brighter; its pulse period is shorter and its profile
narrower, permitting better timing precision; and, most importantly,
it is inclined nearly edge-on to the line of sight from the Earth,
allowing the measurement of Shapiro delay as well as the 3 PK
parameters measurable for PSR~B1913+16.  The orbital parameters for
PSR~B1534+12 are given in Table~\ref{table:params_1534}
\cite{sttw02}.

\begin{table}[h]
\begin{center}
\begin{tabular} {l l}
\hline 
Parameter & Value \\
\hline\hline
Orbital period, $P_b$ (d)   & 0.420737299122(10)  \\ 
Projected semi-major axis, $x$ (s)  & 3.729464(2) \\
Eccentricity, $e$   & 0.2736775(3) \\ 
Longitude of periastron, $\omega$ (deg)  & 274.57679(5)  \\
Epoch of periastron, $T_0$ (MJD)  & 50260.92493075(4) \\
 & \\
Advance of periastron, $\dot\omega$ (deg\,yr$^{-1}$)
  & 1.755789(9) \\
Gravitational redshift, $\gamma$ (ms)   & 2.070(2)  \\
Orbital period derivative, $(\dot P_b)^{\rm obs}$ $(10^{-12})$   &
  $-$0.137(3) \\
Shape of Shapiro delay, $s$ & 0.975(7) \\
Range of Shapiro delay, $r$ ($\mu{\rm s}$)  & 6.7(1.0) \\
\hline
\end{tabular}
\caption{Orbital parameters for PSR~B1534+12 in the DD framework, taken from \cite{sttw02}.\label{table:params_1534}
}
\end{center}
\end{table}

\begin{figure}[h]
  \def\epsfsize#1#2{0.4#1} 
\centerline{\epsfbox{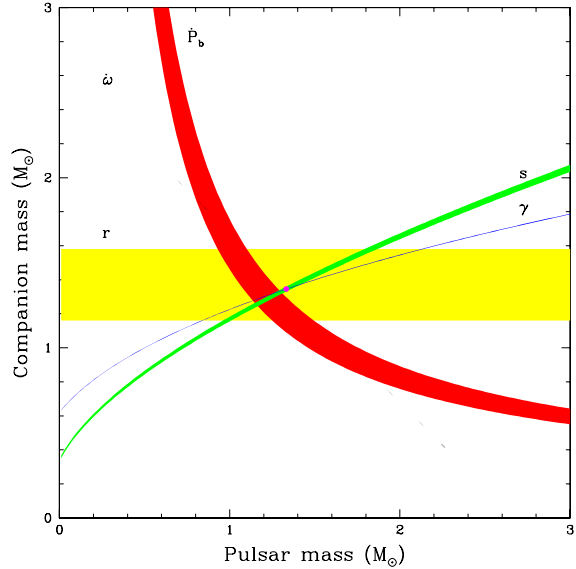}}
  \caption{\it Mass-mass diagram for the PSR B1534+12 system.  Labeled
  curves illustrate 68\% confidence ranges of the DD parameters listed
  in Table~\ref{table:params_1534}.  The filled circle indicates the
  component masses according to the DDGR solution.  The kinematic
  correction for assumed distance $d=0.7\pm0.2\,$kpc has been
  subtracted from the observed value of $\dot{P_b}$; the uncertainty
  on this kinematic correction dominates the uncertainty of this
  curve.  A slightly larger distance removes the small apparent
  discrepancy between the observed and predicted values of
  this parameter.  After \cite{sttw02}.}  \label{figure:m1m2_1534}
\end{figure}

As for PSR~B1913+16, a graphical version of the internal consistency
test is a helpful way to understand the agreement of the measured PK
parameters with the predictions of GR.  This is presented in
Figure~\ref{figure:m1m2_1534}.  It is obvious that the allowed-mass
region derived from the observed value of $\dot P_{\rm b}$ does not in
fact intersect those from the other PK parameters.  This is a
consequence of the proximity of the binary to the Earth, which makes
the ``Shklovskii'' contribution to the observed $\dot P_{\rm b}$ much
larger than for PSR~B1913+16.  The magnitude of this contribution
depends directly on the poorly known distance to the pulsar.  At
present, the best independent measurement of the distance comes from
the pulsar's dispersion measure and a model of the free electron
content of the Galaxy \cite{tc93}, which yield a value of
$0.7\pm0.2$\,kpc.  If GR is the correct theory of gravity, then the
correction derived from this distance is inadequate, and the true
distance can be found by inverting the problem \cite{bb96,sac+98}.
The most recent value of the distance derived in this manner is
$1.02\pm0.05$\,kpc \cite{sttw02}.  (Note that the newer ``NE2001''
Galactic model \cite{cl02} incorporates the GR-derived distance to
this pulsar and hence cannot be used in this case.)  It is possible
that, in the long term, a timing or interferometric parallax may be
found for this pulsar; this would alleviate the $\dot P_{\rm b}$
discrepancy.  The GR-derived distance is in itself interesting, as it
has led to revisions of the predicted merger rate of
double-neutron-star systems visible to gravitational-wave detectors
such as LIGO (e.g., \cite{sac+98,acw99,knst01}) --- although recent
calculations of merger rates determine the most likely merger rates
for particular population models and hence are less vulnerable to
distance uncertainties in any one system \cite{kkl03}.

Despite the problematic correction to $\dot P_{\rm b}$, the other PK
parameters for PSR~B1534+12 are in excellent agreement with each other
and with the values predicted from the DDGR-derived masses.  An
important point is that the three parameters $\dot \omega$, $\gamma$
and $s$ (shape of Shapiro delay) together yield a test of GR to better
than 1\%, and that this particular test incorporates only
``quasi-static'' strong-field effects.  This provides a valuable
complement to the mixed quasi-static and radiative test derived from
PSR~B1913+16, as it separates the two sectors of the theory.

There are three other confirmed double-neutron-star binaries at the
time of writing.  PSR~B2127+11C \cite{akpw90a,and92} is in the
globular cluster M15.  While its orbital period derivative has been
measured \cite{dk96}, this parameter is affected by acceleration in
the cluster potential and the system has not yet proved very useful
for tests of GR, though long-term observations may demonstrate
otherwise.  The two binaries PSRs~J1518+4904 \cite{nst96} and
J1811$-$1736 \cite{lcm+00} have such wide orbits that, although $\dot
\omega$ is measured in each case, prospects for measuring further PK
parameters are dim.  In several circular pulsar--white-dwarf binaries,
one or two PK parameters have been measured --- typically $\dot
\omega$ or the Shapiro delay parameters --- but these do not
over-constrain the unknown masses.  The existing system which provides
the most optimistic outlook is again the pulsar--white-dwarf binary
PSR~J1141-6545 \cite{klm+00a}, for which multiple PK parameters should
be measurable within a few years --- although one may need to consider
the possibility of classical contributions to the measured $\dot
\omega$ from a mass quadrupole of the companion.

\newpage
\subsection{Combined Binary-Pulsar Tests}
\label{section:combined}

\begin{figure}[h]
  \def\epsfsize#1#2{0.4#1}
 \centerline{\epsfbox{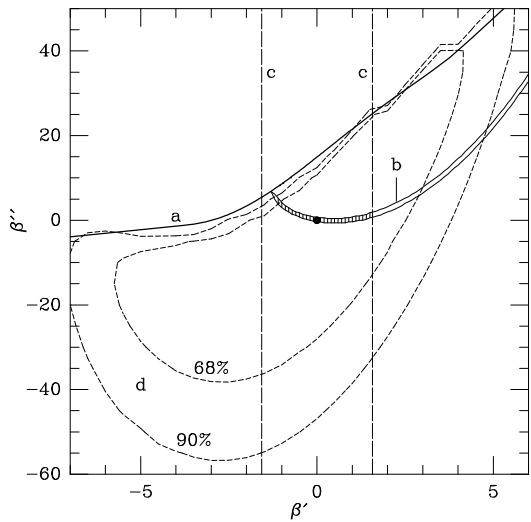}} 
\caption{\it Portions of the tensor--biscalar
$\beta^{\prime}-\beta^{\prime\prime}$ plane permitted by timing
observations of PSRs~B1913+16, B1534+12 and B1855+09 up to
1992. Values lying above the curve labeled ``a'' are incompatible with
the measured $\dot \omega$ and $\gamma$ parameters for PSR~B1913+16.
The curves labeled ``b'' and ``d'' give the allowed ranges of
$\beta^{\prime}$ and $\beta^{\prime\prime}$ for PSRs~B1913+16 and
B1534+12, respectively, fitting for the two neutron-star masses as
well as $\beta^{\prime}$ and $\beta^{\prime\prime}$, using data
available up to 1992.  The vertical lines labeled ``c'' represent
limits on $\beta^{\prime}$ from the SEP-violation test using
PSR~B1855+09 \cite{ds91}. The dot at (0,0) corresponds to GR.
Reprinted by permission from Nature \cite{twdw92}, copyright 1992, 
Macmillan Publishers Ltd.}
\label{figure:twdw92}
\end{figure}

Because of their different orbital parameters and inclinations, the
double-neutron-star systems PSRs~B1913+16 and B1534+12 provide
somewhat different constraints on alternative theories of gravity.
Taken together with some of the limits on SEP violation discussed
above, and with solar-system experiments, they can be used to disallow
certain regions of the parameter space of these alternate theories.
This approach was pioneered by Taylor {\it et al.} \cite{twdw92}, who
combined PK-parameter information from PSRs~B1913+16 and B1534+12 and
the Damour and Sch\"afer result on SEP violation by PSR~B1855+09
\cite{ds91} to set limits on the parameters $\beta^{\prime}$ and
$\beta^{\prime\prime}$ of a class of tensor-biscalar theories
discussed in reference \cite{de92} (Figure~\ref{figure:twdw92}).  In
this class of theories, gravity is mediated by two scalar fields as
well as the standard tensor, but the theories can satisfy the
weak-field solar system tests.  Strong-field deviations from GR would
be expected for non-zero values of $\beta^{\prime}$ and
$\beta^{\prime\prime}$, but the theories approach the limit of GR as
the parameters $\beta^{\prime}$ and $\beta^{\prime\prime}$ approach
zero.

\begin{figure}[h]
  \def\epsfsize#1#2{0.6#1} 
\centerline{\epsfbox{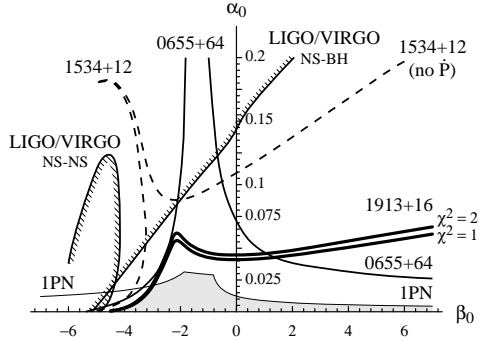}}
  \caption{\it The parameter space in the non-linear $\alpha_0$,
  $\beta_0$ gravitational theory, for neutron stars described by a
  polytrope equation of state.  The regions below the various curves
  are allowed by various pulsar timing limits, by solar-system tests
  (``1PN'') and by projected LIGO/VIRGO observations of NS-NS and
  NS-BH inspiral events.  The shaded region is allowed by all tests.
  The plane and limits are symmetric about $\alpha_0 = 0$.  From
  \cite{de98}; used by permission.}  \label{figure:de98}
\end{figure}

A different class of theories, allowing a non-linear coupling between
matter and a scalar field, was later studied by Damour and
Esposito-Far\`{e}se \cite{de96b,de98}.  The function coupling the
scalar field $\phi$ to matter is given by $A(\phi) =
exp(\frac{1}{2}\beta_0\phi^2)$, and the theories are described by the
parameters $\beta_0$ and $\alpha_0 = \beta_0 \phi_0$, where $\phi_0$
is the value that $\phi$ approaches at spatial infinity
(cf. Section~\ref{section:dipolar}).  These theories allow significant
strong-field effects when $\beta_0$ is negative, even if the
weak-field limit is small.  They are best tested by combining results
from PSRs~B1913+16, B1534+12 (which contributes little to this test),
B0655+64 (limits on dipolar gravitational radiation) and solar-system
experiments (Lunar laser ranging, Shapiro delay measured by Viking
\cite{rsm+79}, and the perihelion advance of Mercury \cite{sha90b}).
The allowed parameter space from the combined tests is shown
graphically in Figure~\ref{figure:de98} \cite{de98}.  Currently, for
most neutron-star equations of state, the solar system tests set a
limit on $\alpha_0$ ($\alpha_0^2 < 10^{-3}$) that is a few times more
stringent than those set by PSRs B1913+16 and B0655+64, although the
pulsar observations do eliminate large negative values of $\beta_0$.
With the limits from the pulsar observations improving only very
slowly with time, it appears that solar-system tests will continue to
set the strongest limits on $\alpha_0$ in this class of theories,
unless a pulsar--black hole system is discovered.  If such a system
were found with a $\sim 10$-$M_{\odot}$ black hole and an orbital
period similar to that of PSR~B1913+16 ($\sim 8$ hours), the limit on
$\alpha_0$ derived from this system would be about 50 times tighter
than that set by current solar-system tests, and 10 times better than
is likely to be achieved by the Gravity Probe B experiment
\cite{de98}.

\newpage
\subsection{Independent Geometrical Information: PSR~J0437$-$4715}
\label{section:psr0437}

\begin{figure}[h]
  \def\epsfsize#1#2{0.4#1} 
 \centerline{\epsfbox{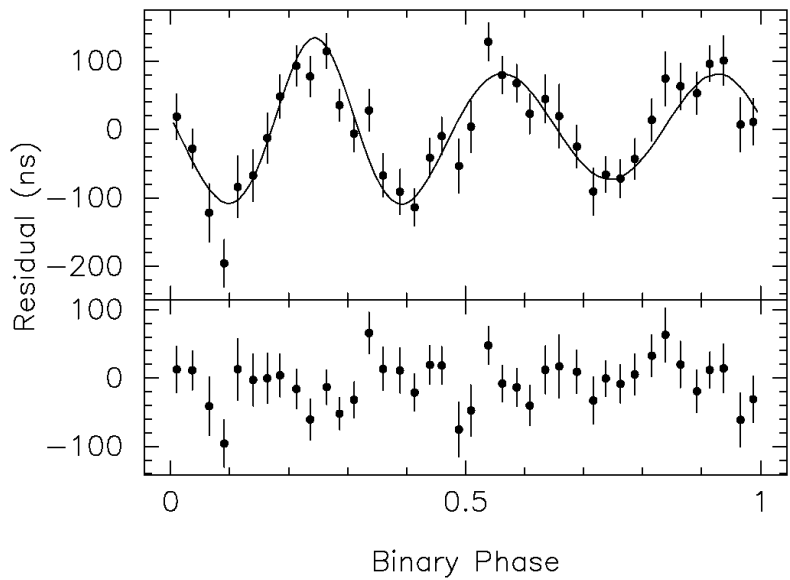}}
  \caption{\it Solid line: predicted value of the Shapiro delay in
  PSR~J0437$-$4715 as a function of orbital phase, based on the
  observed inclination angle of $42^{\circ}\pm9^{\circ}$.  For such
  low-eccentricity binaries, much of the Shapiro delay can be absorbed
  into the orbital Roemer delay; what remains is the $\sim P_{\rm
  b}/3$ periodicity shown.  The points represent the timing residuals
  for the pulsar, binned in orbital phase, and in clear agreement with
  the shape predicted from the inclination angle. Reprinted by permission 
  from Nature \cite{vbb+01}, copyright 2001,
  Macmillan Publishers Ltd.} \label{figure:vbb0437}
\end{figure}

A different and complementary test of GR has recently been permitted
by the millisecond pulsar PSR~J0437$-$4715 \cite{vbb+01}.  At a
distance of only 140\,pc, it is the closest millisecond pulsar to the
Earth \cite{jlh+93}, and is also extremely bright, allowing
root-mean-square timing residuals of 35\,ns with the 64-m Parkes
\cite{parkes} telescope, comparable to or better than the best
millisecond pulsars observed with current instruments at the 300-m
Arecibo telescope \cite{arecibo}.

The proximity of this system means that the orbital motion of the {\it
Earth} changes the apparent inclination angle $i$ of the pulsar orbit
on the sky, an effect known as the annual-orbital parallax
\cite{kop95}.  This results in a periodic change of the projected
semi-major axis $x$ of the pulsar's orbit, written as:
\begin{equation}
x(t) = x_0\left[1+\frac{\cot i}{d}{\bf r}_{\oplus}(t)\cdot{\bf \Omega}^{\prime}\right],
\end{equation}
where ${\bf r}_{\oplus}(t)$ is the time-dependent vector from the
centre of the Earth to the SSB, and ${\bf \Omega}^{\prime}$ is a
vector on the plane of the sky perpendicular to the line of nodes.  A
second contribution to the observed $i$ and hence $x$ comes from the
pulsar system's transverse motion in the plane of the sky
\cite{kop96}:
\begin{equation}
\dot x_{\rm PM} = -x\,\cot i\, {\bf \mu}\cdot{\bf \Omega}^{\prime},
\end{equation}
where ${\bf \mu}$ is the proper motion vector.  By including both
these effects in the model of the pulse arrival times, both the
inclination angle $i$ and the longitude of the ascending node $\Omega$
can be determined \cite{vbb+01}.  As $\sin i$ is equivalent to the
shape of the Shapiro delay in GR (PK parameter $s$), the effect of the
Shapiro delay on the timing residuals can then easily be computed for
a range of possible companion masses (equivalent to the PK parameter
$r$ in GR).  The variations in the timing residuals are well explained
by a companion mass of $0.236\pm0.017\,M_{\odot}$
(Figure~\ref{figure:vbb0437}).  The measured value of $\dot \omega$,
together with $i$, also provides an estimate of the companion mass,
$0.23\pm0.14\,M_{\odot}$, which is consistent with the Shapiro-delay
value.

While this result does not include a true self-consistency check in
the manner of the double-neutron-star tests, it is nevertheless
important, as it represents the only case in which an independent,
purely geometric determination of the inclination angle of a binary
orbit predicts the shape of the Shapiro delay.  It can thus be
considered to provide an independent test of the predictions of GR.

\newpage
\subsection{Spin-Orbit Coupling and Geodetic Precession}
\label{section:geodetic}

\begin{figure}[h]
  \def\epsfsize#1#2{0.4#1}
\centerline{\epsfbox{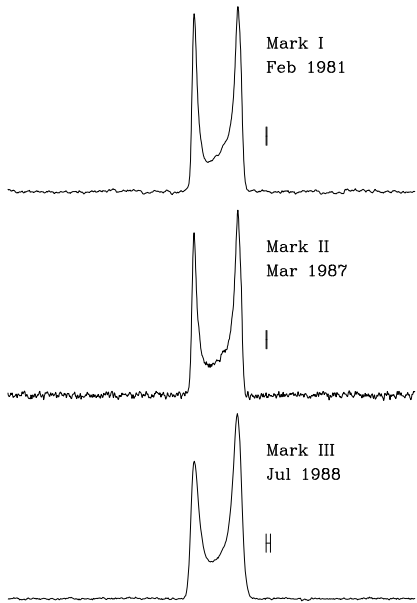}} 
\caption{\it Changes in the observed pulse profile of PSR~B1913+16 throughout 
the 1980s, due to a changing line-of-sight cut through the emission
region of the pulsar.  Taken from reference~\cite{tw89}; used by
permission.}  \label{figure:tw89}
\end{figure}

A complete discussion of GR effects in pulsar observations must
mention geodetic precession, though these results are somewhat
qualitative and do not (yet) provide a model-free test of GR.  In
standard evolutionary scenarios for double-neutron-star binaries
(e.g., \cite{bv91,pk94}), both stellar spins are expected to be
aligned with the orbital angular momentum just before the second
supernova explosion.  After this event, however, the observed pulsar's
spin is likely to be misaligned with the orbital angular momentum, by
an angle of the order of $20^{\circ}$ \cite{bai88}.  A similar
misalignment may be expected when the observed pulsar is the
second-formed degenerate object, as in PSR~J1141$-$6545.  As a result,
both vectors will precess about the total angular momentum of the
system (in practice the total angular momentum is completely dominated
by the orbital angular momentum).  The evolution of the pulsar spin
axis ${\bf S}_1$ can be written as \cite{dr74,bo75}:
\begin{equation} \label{equation:spinevol}
\frac{d{\rm \bf S}_1}{dt} = {\bf \Omega}_1^{\rm spin} \times {\rm \bf
S}_1,
\end{equation}
where the vector ${\bf \Omega}_1^{\rm spin}$ is aligned with the
orbital angular momentum.  Its magnitude is given by:
\begin{equation} \label{equation:omega1spin}
\Omega_1^{\rm spin} = \frac{1}{2}\left(\frac{P_{\rm b}}{2 \pi}\right)^{-5/3}
\frac{m_2(4m_1+3m_2)}{(1-e^2)(m_1+m_2)^{4/3}}T_{\odot}^{2/3},
\end{equation}
where $T_\odot\equiv GM_\odot/c^3 = 4.925490947\,\mu$s, as in
Section~\ref{section:pkparms}.  This predicted rate of precession is
small; the three systems with the highest $\Omega_1^{\rm spin}$ values
are: PSR~J1141$-$6545 at $1.35^{\circ}$\,yr$^{-1}$, PSR~B1913+16 at
$1.21^{\circ}$\,yr$^{-1}$ and PSR~B1534+12 at
$0.52^{\circ}$\,yr$^{-1}$.  The primary manifestation of this
precession is expected to be a slow change in the shape of the pulse
profile, as different regions of the pulse emission beam move into the
observable region.

\begin{figure}[t]
  \def\epsfsize#1#2{0.4#1}
  \centerline{\epsfbox{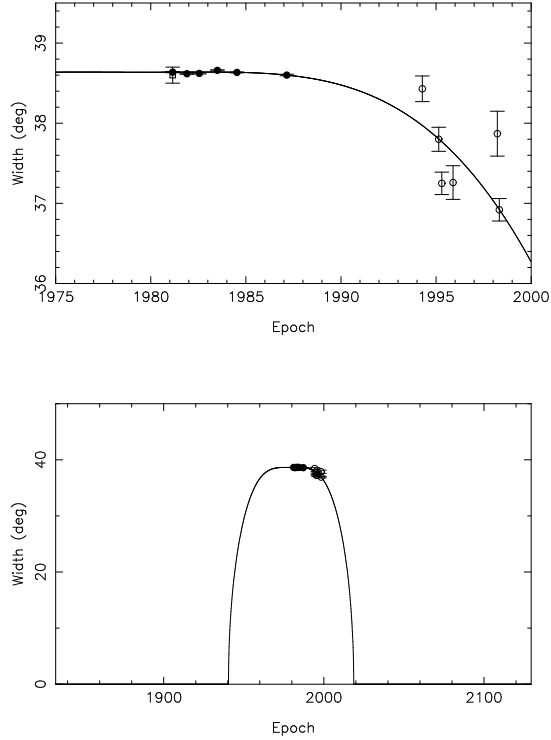}} 
\caption{\it Top: change in peak
  separation of the relativistic double-neutron-star binary
  PSR~B1913+16, as observed with the Arecibo (solid points,
  \cite{wrt89}) and Effelsberg (open circles, \cite{kra98})
  telescopes.  Bottom: projected disappearance of PSR~B1913+16 in
  approximately 2025.  Taken from reference~\cite{kra98}; used by
  permission.}  \label{figure:kra98}
\end{figure}

\begin{figure}[t]
  \def\epsfsize#1#2{0.4#1}
  \centerline{\epsfbox{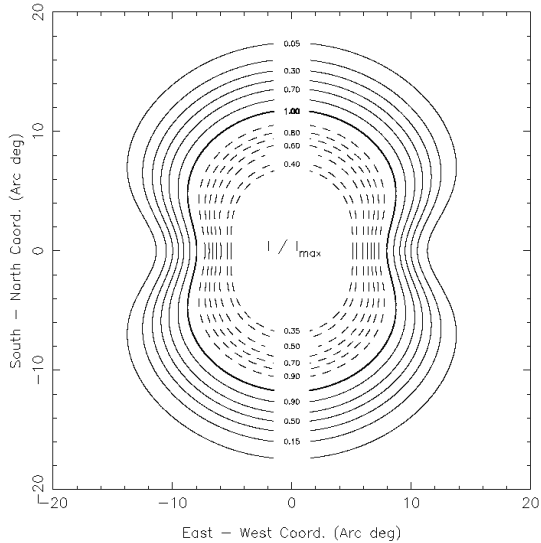}} 
\caption{\it Hourglass-shaped beam for PSR~B1913+16 derived from the 
symmetric-component analysis of \cite{wt02}.  Taken from
reference~\cite{wt02}; used by permission.}  \label{figure:wt02}
\end{figure}

\begin{figure}[t]
  \def\epsfsize#1#2{0.4#1}
  \centerline{\epsfbox{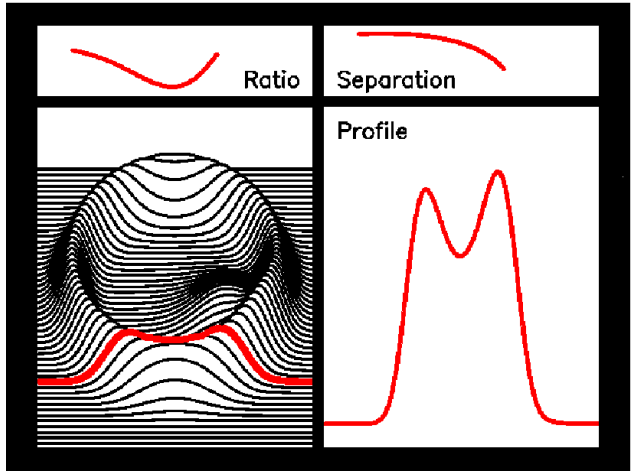}} 
\caption{\it Alternate proposed beam shape for PSR~B1913+16, consisting of a 
symmetric cone plus an offset core.  The red lines indicate an
example cut through the emission region, as well as the predicted
pulse peak ratio and separation as functions of time. After~\cite{kra02}, 
courtesy Michael Kramer.}
\label{figure:kra_1913}
\end{figure}

Evidence for long-term profile shape changes is in fact seen in
PSRs~B1913+16 and B1534+12.  For PSR~B1913+16, profile shape changes
were first reported in the 1980s \cite{wrt89}, with a clear change
in the relative heights of the two profile peaks over several years
(Figure~\ref{figure:tw89}).  No similar changes were found in the
polarization of the pulsar \cite{cwb90}.  Interestingly, although a
simple picture of a cone-shaped beam might lead to an expectation of a
change in the {\it separation} of the peaks with time, no evidence for
this was seen until the late 1990s, at the Effelsberg 100-m telescope
\cite{kra98}, by which point the two peaks had begun to move closer
together at a rather fast rate.  Kramer \cite{kra98} used this
changing peak separation, along with the predicted precession rate and
a simple conal model of the pulse beam, to estimate a spin-orbit
misalignment angle of about $22^{\circ}$ and to predict that the
pulsar will disappear from view in about 2025 (see
Figure~\ref{figure:kra98}), in good agreement with an earlier
prediction by Istomin \cite{ist91} made before the peak separation
began to change.  Recent results from Arecibo \cite{wt02} confirm the
gist of Kramer's results, with a misalignment angle of about
$21^{\circ}$.  Both sets of authors find there are four degenerate
solutions that can fit the profile separation data; two can be
discarded as they predict an unreasonably large misalignment angle of
$\sim 180^{\circ}- 22^{\circ} = 158^{\circ}$ \cite{bai88}, and a third
is eliminated because it predicts the wrong direction of the position
angle swing under the Rotating Vector Model \cite{rc69a}.  The main
area of dispute is the actual shape of the emission region; while
Weisberg and Taylor find an hourglass-shaped beam (see
Figure~\ref{figure:wt02}), Kramer maintains that a nearly circular
cone plus an offset core is adequate (see Figure~\ref{figure:kra_1913}).
In any event, it is clear that the interpretation of the profile
changes requires some kind of model of the beam shape.  Kramer
\cite{kra02,klk03} lets the rate of precession vary as another free
parameter in the pulse-shape fit, and finds a value of
$1.2^{\circ}\pm0.2^{\circ}$.  This is consistent with the GR
prediction but still depends on the beam-shape model and is therefore
not a true test of the precession rate.

\begin{figure}[h]
  \def\epsfsize#1#2{0.4#1} 
\centerline{\epsfbox{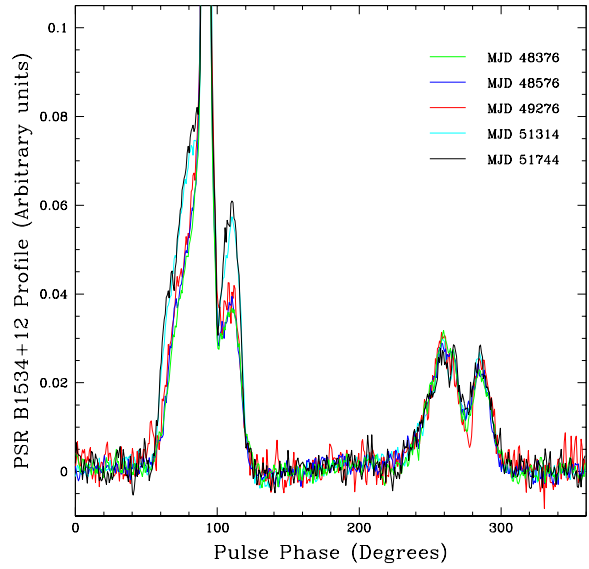}}
  \caption{\it Evolution of the low-level emission surrounding the
  main pulse of PSR~B1534+12, over a period of nearly 10 years, as
  measured with the Arecibo telescope \cite{arecibo}.  Stairs et al.,
  unpublished.}  \label{figure:1534evol}
\end{figure}

PSR~B1534+12, despite the disadvantages of a more recent discovery and
a much longer precession period, also provides clear evidence of
long-term profile shape changes.  These were first noticed at
1400\,MHz by Arzoumanian \cite{arz95,atw99} and have become more
obvious at this frequency and at 430\,MHz in the post-upgrade period
at Arecibo \cite{stta00}.  The principal effect is a change in the
low-level emission near to the main pulse
(Figure~\ref{figure:1534evol}), though related changes in polarization
are now also seen.  As this pulsar shows polarized emission through
most of its pulse period, it should be possible to form a better
picture of the overall geometry than for PSR~B1913+16; this may make
it easier to derive an accurate model of the pulse beam shape.

As for other tests of GR, the pulsar--white-dwarf binary
PSR~J1141$-$6545 promises interesting results.  As noted by the
discoverers \cite{klm+00a}, the region of sky containing this pulsar
had been observed at the same frequency in an earlier survey
\cite{jlm+92}, but the pulsar was not seen, even though it is now
very strong.  It is possible that interference corrupted that original
survey pointing, or that a software error prevented its detection, but
it is also plausible that the observed pulsar beam is evolving so
rapidly that the visible beam precessed into view during the 1990s.
Clearly, careful monitoring of this pulsar's profile is in order.

\newpage
\section{Conclusions and Future Prospects}
\label{section:conclusions}

The tremendous success to date of pulsars in testing different aspects
of gravitational theory leads naturally to the question of what can be
expected in the future.  Improvements to the equivalence-principle
violation tests will come from both refining the timing parameters of
known pulsars (in particular, limits on eccentricities and orbital
period derivatives) and the discovery of further pulsar--white-dwarf
systems.  Potentially coalescing pulsar--white-dwarf binaries, such as
PSRs~J1141$-$6545, J0751+1807 \cite{lzc95} and 1757$-$5322
\cite{eb01a}, bear watching from the point of view of limits on
dipolar gravitational radiation.  Another worthy, though difficult,
goal is to attempt to derive the full orbital geometry for
ultra-low-eccentricity systems, as has been done for PSR~J0437$-$4715
\cite{vbb+01}; this would quickly lead to significant improvements
in the eccentricity-dependent tests.

The orbital-period-derivative measurements of double-neutron-star
binaries are already limited more by systematics (Galactic
acceleration models for PSR~B1913+16, poorly known distance for
PSR~B1534+12) than by pulsar timing precision.  However, with improved
Galactic modeling and a realistic expectation of an interferometric
(VLBI) parallax for PSR~B1534+12, there is still hope for testing more
carefully the prediction of quadrupolar gravitational radiation from
these systems.  The other timing parameters, equally important for
tests of the quasi-static regime, can be expected to improve with time
and better instrumentation, such as the wider-bandwidth coherent
dedispersion systems now being installed at many observatories
(e.g.,\cite{cobra,cpsr2}).  Especially exciting would be a
measurement of the elusive Shapiro delay in PSR~B1913+16; the
longitude of periastron is now precessing into an angular range where
it may facilitate such a measurement \cite{wt03}.

In the last few years, surveys of the Galactic Plane and flanking
regions, using the 64-m Parkes telescope in Australia \cite{parkes},
have discovered several hundred new pulsars (e.g.,
\cite{mlc+01,ebvb01}), including several new circular-orbit
pulsar--white-dwarf systems \cite{eb01a,eb01b,clm+01} and the
eccentric pulsar--white-dwarf binary PSR~J1141$-$6545 \cite{klm+00a}.
A complete reprocessing of the Galactic Plane survey with improved
interference filtering is in progress; thus there is still hope that a
truly new system such as a pulsar--black-hole binary may emerge from
this large survey.  Several ongoing smaller surveys of small regions
and globular clusters (e.g., \cite{clf+00,rhs+03}) are also finding a
number of new and exotic binaries, some of which may eventually turn
out to be useful for tests of GR.  The possible recent appearance of
PSR~J1141$-$6545 and the predicted disappearance of PSR~B1913+16 due
to geodetic precession make it worthwhile to periodically revisit
previously surveyed parts of the sky in order to check for
newly-visible exotic binaries.  Over the next several years,
large-scale surveys are planned at Arecibo \cite{alfa} and the new
100-m Green Bank Telescope \cite{gbt}, offering the promise of over
1000 new pulsars including interesting binary systems.  The
sensitivity of these surveys will of course be dwarfed by the
potential of the proposed Square Kilometre Array radio telescope
\cite{ska}, which will be sensitive to pulsars clear through our
Galaxy and into neighbouring galaxies such as M31.  The next decade or
two promise to be exciting times for pulsar searchers and for those
looking to set ever-more-stringent limits on deviations from General
Relativity.

\newpage



\section{Acknowledgements}
\label{section:acknowledgements}

The author holds an NSERC University Faculty Award and is supported by
a Discovery Grant.  She thanks Michael Kramer, George Hobbs and Zaven
Arzoumanian for careful readings of the manuscript, Duncan Lorimer for
generously sharing his keyworded reference list, and Gilles
Esposito-Far{\`e}se, Michael Kramer, Joe Taylor, Steve Thorsett,
Willem van Straten and Joel Weisberg for allowing reproduction of
figures from their work.  The Arecibo Observatory, a facility of the
National Astronomy and Ionosphere Center, is operated by Cornell
University under a cooperative agreement with the National Science
Foundation.  The National Radio Astronomy Observatory is a facility of
the National Science Foundation operated under cooperative agreement
by Associated Universities, Inc.  The Parkes radio telescope is part
of the Australia Telescope which is funded by the Commonwealth of
Australia for operation as a National Facility managed by CSIRO.

\newpage


\end{document}